\documentclass[%
reprint,
showpacs,
 amsmath,amssymb,aps,
prb,
]{revtex4-1}

\usepackage{graphicx}

\begin{document}

\preprint{APS/123-QED}

\title{Magnetic phase diagram and electronic structure of UPt$_2$Si$_2$ at high magnetic fields: a possible field-induced Lifshitz transition}

\author{D. Schulze Grachtrup$^{1}$, N. Steinki$^{1}$, S. S\"ullow$^{1}$, Z. Cakir$^{2}$, G. Zwicknagl$^{2}$, Y. Krupko$^{3}$, I. Sheikin$^{3}$, M. Jaime$^{4}$, J.A. Mydosh$^{5}$}

\address{
$^{1}$Institut f\"ur Physik der Kondensierten Materie, TU Braunschweig, D-38106 Braunschweig, Germany\\
$^{2}$Institut f\"ur Mathematische Physik, TU Braunschweig, D-38106 Braunschweig, Germany\\
$^{3}$Laboratoire National des Champs Magn$\acute{\textrm{e}}$tiques Intenses (LNCMI-EMFL), CNRS, UGA, 38042 Grenoble, France\\
$^{4}$National High Magnetic Field Laboratory, Los Alamos National Laboratory, Los Alamos, New Mexico 87545, USA\\
$^{5}$Kamerlingh Onnes Laboratory and Institute Lorentz, Leiden University, 2300RA Leiden, The Netherlands}

\date{\today}

\begin{abstract}
We have measured Hall effect, magnetotransport and magnetostriction on the field induced phases of single crystalline UPt$_2$Si$_2$ in magnetic fields up to 60\,T at temperatures down to 50\,mK, this way firmly establishing the phase diagram for magnetic fields $B \| a$ and $c$ axes. Moreover, for $B \| c$ axis we observe strong changes in the Hall effect at the phase boundaries. From a comparison to band structure calculations utilizing the concept of a dual nature of the uranium 5$f$ electrons, we propose that these represent field induced topological changes of the Fermi surface due to at least one Lifshitz transition. Furthermore, we find a unique history dependence of the magnetotransport and magnetostriction data, indicating that the proposed Lifshitz type transition is of a discontinuous nature, as predicted for interacting electron systems.
\end{abstract}

\maketitle

\section{Introduction}

Lifshitz transitions - that is, quantum phase transitions involving topological changes of the Fermi surface, and thus referred to as electronic topological transitions (ETT) - have been proposed to play a major role for the physics of correlated electron systems. Here, a variety of exotic field-induced phases, as well as unconventional pressure induced phases (including unconventional superconducting ones) have been observed and attributed to Lifshitz transitions \cite{kozlova,rourke,plackowski,yelland,deppe,Pfau13,Pourret13,Naren13,aoki14,aoki16,bastien16}. 

The theory of ETT was developed to account for the ground state properties of certain materials under wide variation of external parameters such as pressure \cite{lifshitz,Varlamov1989}. It considered non-interacting electrons at zero temperature, yielding a continuous transition of 2$\frac{1}{2}$ order, which reflects the exponent in the Ehrenfest expression in three dimensions. Later, based on various experimental observations, the case of interacting electrons was treated in detail \cite{yamaji2006,yamaji2007,schlottmann,bercx,wang}. Here, conceptually, a new (low) energy scale is associated with the interacting electron system, which may produce ETT in experimentally accessible magnetic field and pressure ranges of a few 10\,T and GPa. As well, it was predicted that for interacting electron systems the transitions inherently become discontinuous \cite{yamaji2006,yamaji2007}. 

Regarding the experimental verification of electronic topological transitions, cases of real materials exhibiting Lifshitz transitions are rare. On general grounds, it has been demonstrated that anomalies from ETT should be observable in various transport properties \cite{Varlamov1989,Sharapov03}. Yet, ETT exist only for zero temperature, and smear out with finite temperature. It is a formidable experimental task to identify a Lifshitz type transition, requiring experiments down to low temperatures under extreme conditions. As well, for correlated electron systems, calculating the band structure as function of external control parameters is a very challenging task.

A case in point is the intermetallic $5f$ electron system UPt$_2$Si$_2$. The material belongs to the large class of U$T_2M_2$-compounds ($T$ = transition metal, $M$ = Si or Ge) and crystallizes in the tetragonal CaBe$_2$Ge$_2$ structure (space group P4/{\it nmm}) \cite{hiebl1987}. In zero magnetic field, it undergoes an antiferromagnetic (AFM) transition at $T_N = 32$\,K. The magnetic structure consists of moments $\mu_{ord} \sim 2.5 \mu_B$ ferromagnetically aligned within the $ab$ plane, and antiferromagnetically coupled and pointing along the $c$ axis \cite{steeman,sullow4}. The simple magnetic structure with a large magnetic moment, combined with a moderately enhanced electronic contribution to the specific heat $\gamma = 32$\,mJ/mole\,K$^2$, was taken as indicator for UPt$_2$Si$_2$ to be one of the rare examples of an uranium intermetallic local moment magnet. Correspondingly, a crystal electric field (CEF) scheme for the $5f^2$ state of the uranium ion was proposed, that has been used to explain initial high field magnetization measurements and the anisotropy of the susceptibility \cite{nieuwenhuys,amitsuka2}. Additional fine structure in the magnetization observable in the field range $\sim 20 - 40$\,T was not considered to be at odds with the CEF concept.

Based on an extensive reinvestigation of the magnetization we have demonstrated that the agreement between CEF model and experimental data does not hold up to high fields. For magnetic fields $B \parallel a$ axis, aside from the suppression of AFM order, there is a hysteretic high field ($\sim$\,40\,T) regime, whose nature is not understood as yet \cite{schulze2,schulze} (see Fig.~\ref{fig:pd}). Moreover, for fields $B \parallel c$ axis, above 24\,T the experimental data strongly deviates from the CEF theoretical predictions. In particular, a complex series of field induced phases is observed above 24\,T, and which cannot be attributed only to spin reorientation processes and/or crystal field effects \cite{schulze,schulze2,nieuwenhuys,amitsuka2}. As an alternative to the CEF model, we have proposed that an itinerant picture of the properties of UPt$_2$Si$_2$ is more appropriate, a view supported by recent band structure calculations \cite{Elgazzar12}. Moreover, these calculations have highlighted the relevance of correlation effects in this system \cite{Elgazzar12,zwicknagl16}. As well, the general character of the band structure has been revealed to be ''quasi-two-dimensional'' as result of the comparatively low crystallographic symmetry. This two-dimensional character is reflected for instance in the highly anisotropic resistivity of UPt$_2$Si$_2$ \cite{sullow4}.

In Ref.~\cite{schulze2}, we have argued that the observation of the field induced phases in UPt$_2$Si$_2$ is related to Lifshitz type transitions. To test the validity of this concept in UPt$_2$Si$_2$, studies at lowest temperatures and using experimental tools directly testing the Fermi surface (FS) and the order of the phase transitions are required. A test of the FS by means of quantum oscillation measurements cannot be performed for UPt$_2$Si$_2$, as the oscillations are suppressed by intrinsic structural disorder from strained Pt(2)/Si(2) layers in the CaBe$_2$Ge$_2$ lattice \cite{sullow4}. Therefore, more integral -- and less disorder--dependent -- probes of the FS need to be investigated to check the Lifshitz scenario for UPt$_2$Si$_2$. In addition, to establish the order of the field-induced phase transitions experimental probes sensitive to the structural properties may be used. If combined with band structure calculations, it will allow an assessment of the nature of the field-induced phases in UPt$_2$Si$_2$.  

In this situation, we present a study on UPt$_2$Si$_2$ under extreme conditions, that is at temperatures down to 50\,mK and in fields up to 60\,T, using Hall effect, magnetotransport and magnetostriction. Our experiments clearly demonstrate changes of the FS in high magnetic fields. Our study is complemented by band structure calculations utilizing the concept of a dual nature of the uranium 5$f$ electrons, which simulate the effect of magnetic fields on the topology of the band structure. Resulting from these calculations, it is verified that Lifshitz type transitions may be induced in high magnetic fields in UPt$_2$Si$_2$. 

\section{Experimental details}

The experiments presented here were performed on single crystalline UPt$_2$Si$_2$, the samples being as-cast, bar shaped with a cross section $\sim 1 \times 1$\,mm$^2$ and length of a few mm. Material from the same single crystal has been characterized in the Refs.~[\onlinecite{schulze,schulze2,sullow4,bleckmann}]. 

The electronic transport studies were carried out in the Laboratoire National des Champs Magn$\acute{\textrm{e}}$tiques Intenses in dc fields up to 34\,T directed along $a$ and $c$ axes. For the experiments inside the magnet bore a dilution cryostat was installed. Data were taken with a standard lock-in setup, with a reasonable signal-to-noise ratio obtained with a measurement current of 1\,mA directed along the $a$ axis. It resulted in an equilibrium temperature of 120\,mK, with additional experiments with lower currents carried out down to 50\,mK. Accordingly, we now have access to a wide range of the field/temperature plane up to a $B/T$ ratio of almost 700 T/K. 

The sample was fitted onto a rotatable sample holder and immersed into the helium mixture. Up to 10 electrical contacts were glued on the sample surfaces with silver paint to allow for simultaneaous measurements of transverse magnetoresisitivity and Hall effect for each magnetic field direction. In a second round of experiments, the same set-up was used, but now to measure the longitudinal magnetoresistivity. Finally, axial magnetostriction experiments were performed at the Los Alamos High Field Laboratory in pulsed fields up to 60\,T directed along $a$ and $c$ axes. Here, the base temperature was 1.4\,K, with the experiment performed using an optical fiber with a Bragg grating \cite{daou,jaime}. 

\section{Results}

\begin{figure}[htb]
\centering
\includegraphics[width=1 \linewidth]{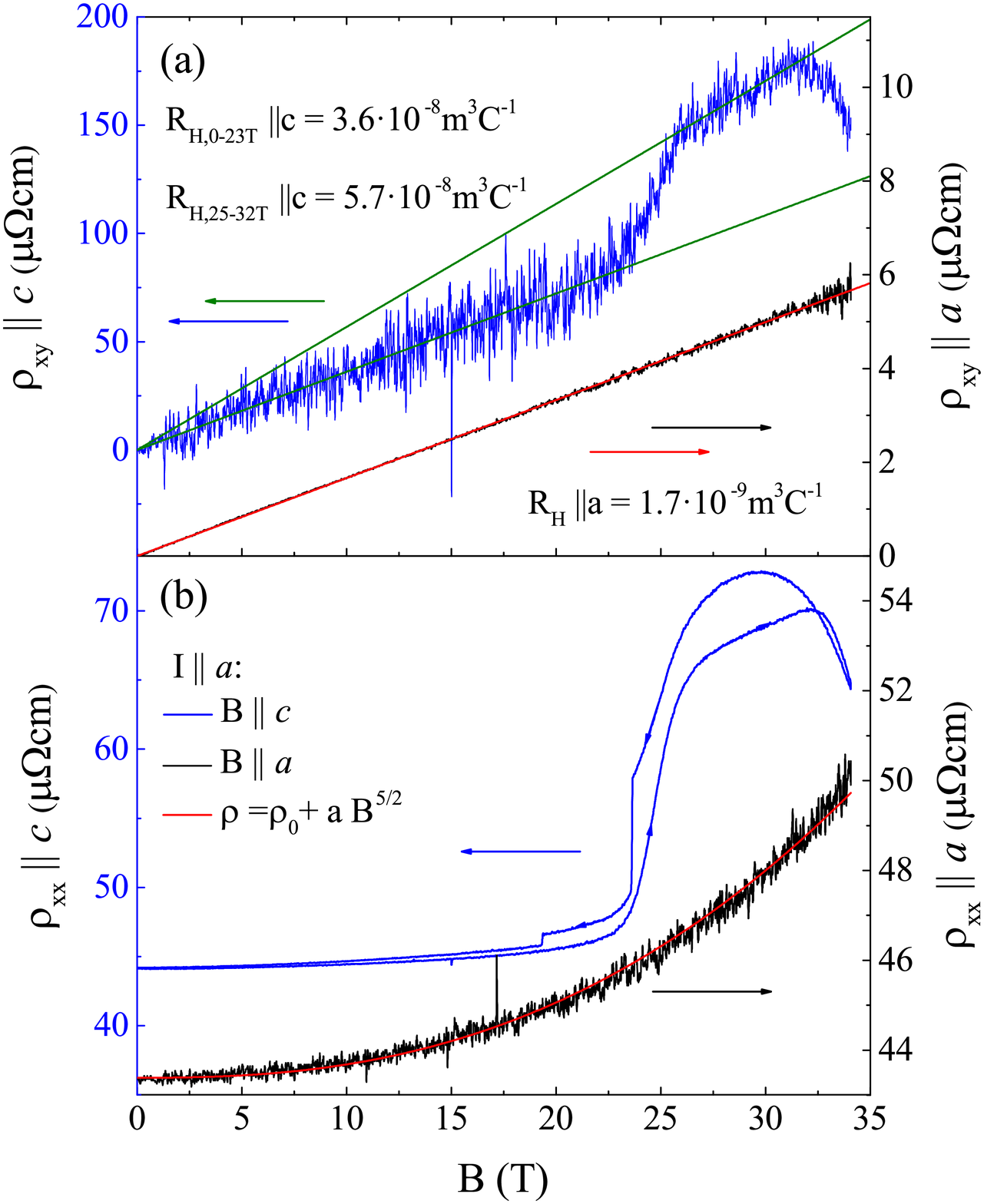} \\
\caption{(Color online) Comparative plot of a.) Hall resistivity $\rho_{xy}$ at $T = 300$\,mK for $B \parallel a$ and 130\,mK for $B \parallel c$ axis, together with fits to the data, and b.) transverse magnetoresistivity of UPt$_2$Si$_2$ at $T = 300$\,mK for $B \parallel a$ and 120\,mK for $B \parallel c$; for details see text.}
\label{fig:hall}
\end{figure}

\begin{figure}[htb]
\centering
\includegraphics[width=1 \columnwidth]{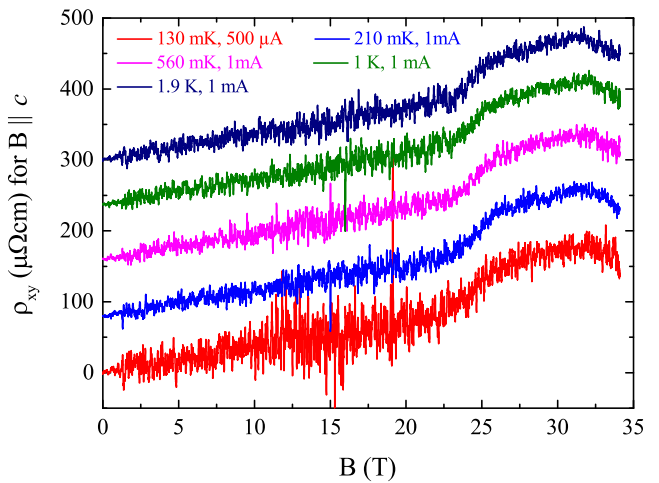}
\caption{(Color online) Hall resistivity $\rho_{xy}$ at various temperatures for $B \parallel c$ axis, data shifted for clarity; for details see text.}
\label{fig:halls}
\end{figure}

For fields $B \parallel a$ axis up to 34\,T and low $T$, the Hall resistivity $\rho_{xy}$ is linear in $B$ (Fig.~\ref{fig:hall}(a)). Consistent with Ref.~[\onlinecite{schulze2}], there are no phase transitions in this field range. Next, Hall effect data for $B \parallel c$ axis up to 34\,T are included in Fig.~\ref{fig:hall}(a), the full set of (low temperature) data is summarized in Fig.~\ref{fig:halls}. For fields up to 23\,T, {\it viz.}, in the AFM phase $I$, $\rho_{xy}$ increases linearily with field~\cite{schulze2}. At the phase boundary $I - III$ there is a distinct upward curvature in $\rho_{xy}(B)$, which becomes linear in field again in phase $III$. At the phase $III - V$ boundary there is a - now downwards - curvature in $\rho_{xy}$. Within experimental scatter, no hysteresis is observed between field-sweep-up and -down measurements, and there is no temperature dependence in the range below 2\,K. Qualitatively, this behavior is reminiscent to that of Rh-doped URu$_2$Si$_2$ \cite{oh}. 

For parameterization, the Hall effect is fitted using the expression $\rho_{xy} = R_H B$. For $B \parallel a$ axis the data in phase $I$ yield a Hall coefficient $R_H = 1.7 \times 10^{-9}$\,m$^3$/C (solid line in Fig.~\ref{fig:hall}(a)). Correspondingly, for $B \parallel c$ axis the linear regimes from 0 to 23\,T and 25 to 32\,T lead to Hall coefficients $R_H = 3.6 \times 10^{-8}$\,m$^3$/C and $R_H = 5.7 \times 10^{-8}$\,m$^3$/C, respectively. Overall, these values are broadly consistent with the typical behavior of heavy fermion intermetallics. We note that, although for $B \parallel c$ axis at highest fields (34\,T) the system resides in phase $V$ (as is proven by the observation of hysteresis in the magnetoresistivity, see below), in the Hall effect we observe non-monotonic behavior in this field range. The reason for the unusual behavior is not clear, and will require experiments to still higher fields to solve.

\begin{figure}[htb]
\centering
\includegraphics[width=1 \columnwidth]{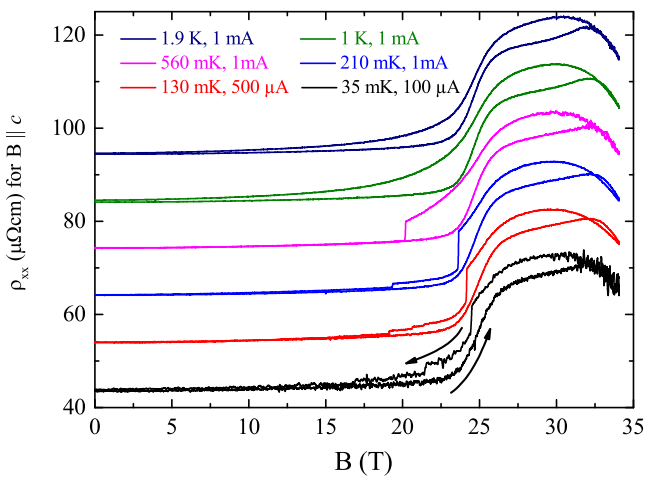}
\caption{(Color online) Transverse magnetoresistivity of UPt$_2$Si$_2$ at various temperatures for $B \parallel c$ axis; for details see text.}
\label{fig:MRs}
\end{figure}

In magnetic materials, the Hall effect contains two terms. The {\it normal} contribution $\rho_{xy}^{nor} = R_N B$ measures the carrier density $n$ in units of the electron charge $e$: $R_N = \left( n e \right)^{-1}$. The {\it anomalous} Hall contribution $\rho_{xy}^{ano}$ reflects terms dependent on the resistivity and/or magnetization (reported in the Refs.~[\onlinecite{schulze,schulze2,amitsuka2}]). Therefore, and adding to the data published in Ref.~[\onlinecite{schulze2}], low temperature magnetoresistivity has been carried out (Figs.~\ref{fig:hall}(b) and \ref{fig:MRs}). At lowest temperatures for $B \parallel a$ axis up to 34\,T, for the magnetoresistivity we find to good approximation $\rho_{xx}(B) = \rho_{xx}(B=0) + a B^{\frac{5}{2}}$. In accordance with the Hall effect and Ref.~[\onlinecite{schulze2}], we find no evidence for phase transitions.

In contrast, for $B \parallel c$ axis the transitions from phase $I$ into $III$ and $III$ into $V$ are reflected by distinct anomalies in the magnetoresistivity. The transition $I \rightarrow III$ is accompanied by a steep increase of the magnetoresistivity, with the midpoint of the upturn close to the transition field determined from magnetization. Conversely, the transition $III \rightarrow V$ shows up as a corresponding drop of the magnetoresistivity.

\begin{figure}[htb]
\centering
\includegraphics[width=1 \columnwidth]{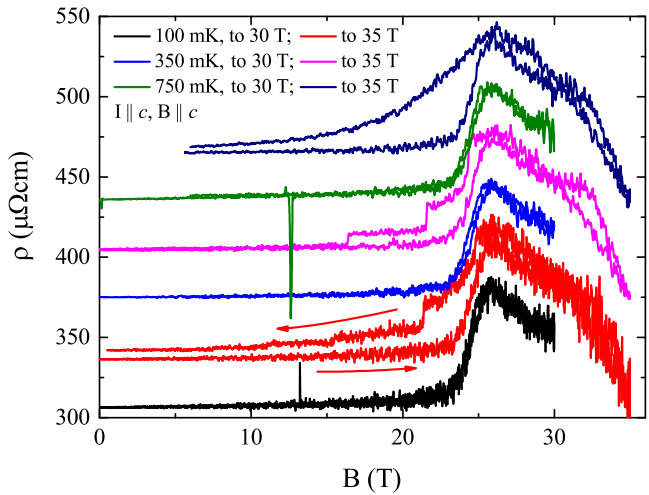}
\caption{(Color online) Longitudinal magnetoresistivity of UPt$_2$Si$_2$ at various temperatures for $B \parallel c$ axis for field sweeps either into phase $III$ (30\,T) and back to zero field, or into phase $V$ (35\,T) and back.}
\label{fig:lmrs}
\end{figure}

Surprisingly, the magnetoresistivity $B \parallel c$ axis is accompanied by a curious type of hysteresis (Figs.~\ref{fig:hall}(b), \ref{fig:MRs}, \ref{fig:lmrs}): Measurements of $\rho_{xx}(B)$ by sweeping from zero field into phase $V$ and back produce hysteresis in the magnetoresistivity. In contrast, sweeps from zero field only into phase $III$ and back produce no hysteresis in $\rho_{xx}(B)$ at low $T$. To demonstrate this, we have carried out field history dependent longitudinal magnetoresistivity measurements ($B \parallel c \parallel I$) at low temperatures, which we summarize in Fig.~\ref{fig:lmrs}. Here, we have first swept the field from zero into phase $III$ (up to 30\,T) and back to zero for various temperatures below 1\,K. For this measurement sequence no hysteresis is observed in the magnetoresistivity. Conversely, for field sweeps at the same temperatures up into phase $V$ (final field: 35\,T) we detect hysteresis in the magnetoresistivity in field-sweep-down vs. field-sweep-up data. This observation indicates that the phase $III-V$ boundary denotes a first order phase transition. Moreover, this observation is consistent with our previous magnetoresistivity study \cite{schulze2}, where we swept the field up to 28\,T, {\it i.e.}, into phase $III$, and did not observe hysteresis.

In terms of the anomalous Hall contribution $\rho_{xy}^{ano}$, the absence of hysteresis in the Hall effect and its presence in the magnetoresistivity (Fig.~\ref{fig:hall}) implies that $\rho_{xy}^{ano}$ is not dependent on $\rho_{xx}(B)$. Then, the upturn in $\rho_{xy}$ at the phase $I-III$ boundary might be attributed to the corresponding upturn in the magnetization $M$ (see Refs. [\onlinecite{schulze,schulze2,amitsuka2}]). Conversely, at the phase $III-V$ boundary the downturn in $\rho_{xy}$ is clearly at odds with the upturn in $M$. It implies that this phase transition is accompanied by a carrier density change, and which may involve a qualitative change of the Fermi surface as in an ETT.

\begin{figure}[htb]
\includegraphics[width=1 \linewidth]{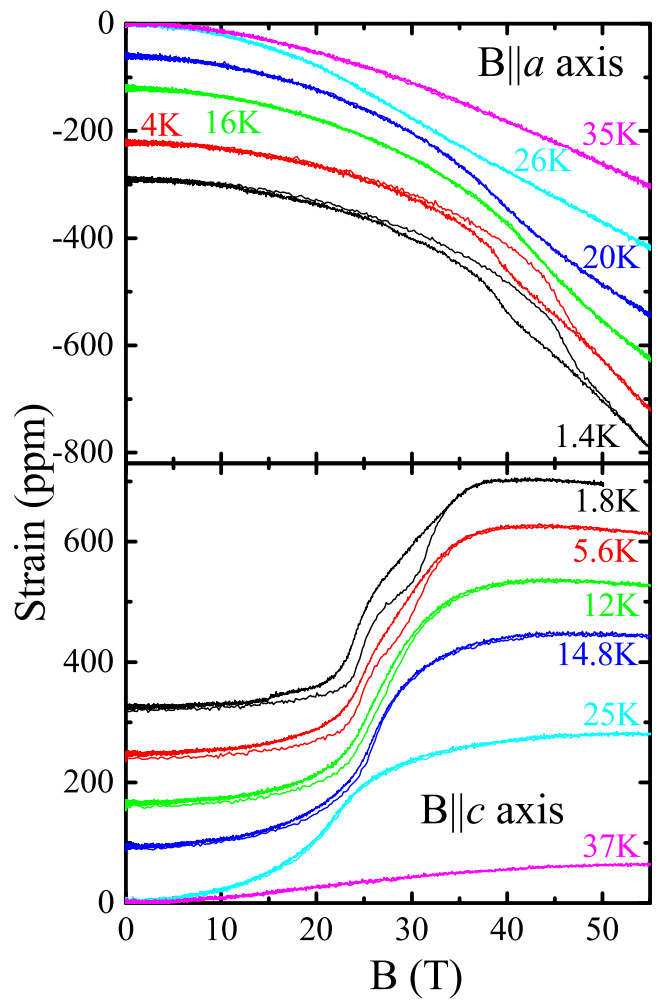}
\caption{(Color online) Axial magnetostriction of UPt$_2$Si$_2$ for magnetic fields $B \parallel a$ and $c$ axes plotted up to 55\,T for various temperatures; data shifted for clarity, for details see text.}
\label{fig:MS}
\end{figure}

To complement our study on UPt$_2$Si$_2$ with a structural probe, we have carried out axial magnetostriction experiments, which we depict in Fig.~\ref{fig:MS}. For $B \parallel a$ axis we find a contraction of the sample for all fields. Further, a slight change of slope occurs at elevated fields, becoming hysteretic in the temperature/field range, where magnetization hysteresis is observed. The (hysteretic) features denote the transition from phase $I$ into the paramagnetic state. 

For $B \parallel c$ axis the crystal UPt$_2$Si$_2$ expands for fields up to $\sim$\,40\,T, and at highest fields the magnetostriction saturates. The field-induced phase transitions are identified as additional structure in the magnetostriction. Similar to the magnetization, at low temperatures there is a twofold structure in the data reflecting the transition from phase $I$ into $III$, and finally into phase $V$. The critical fields of the different phases are identified as points of maximum slope in the field-sweep-up measurements. 
 
\begin{figure}[htb]
\includegraphics[width=1 \linewidth]{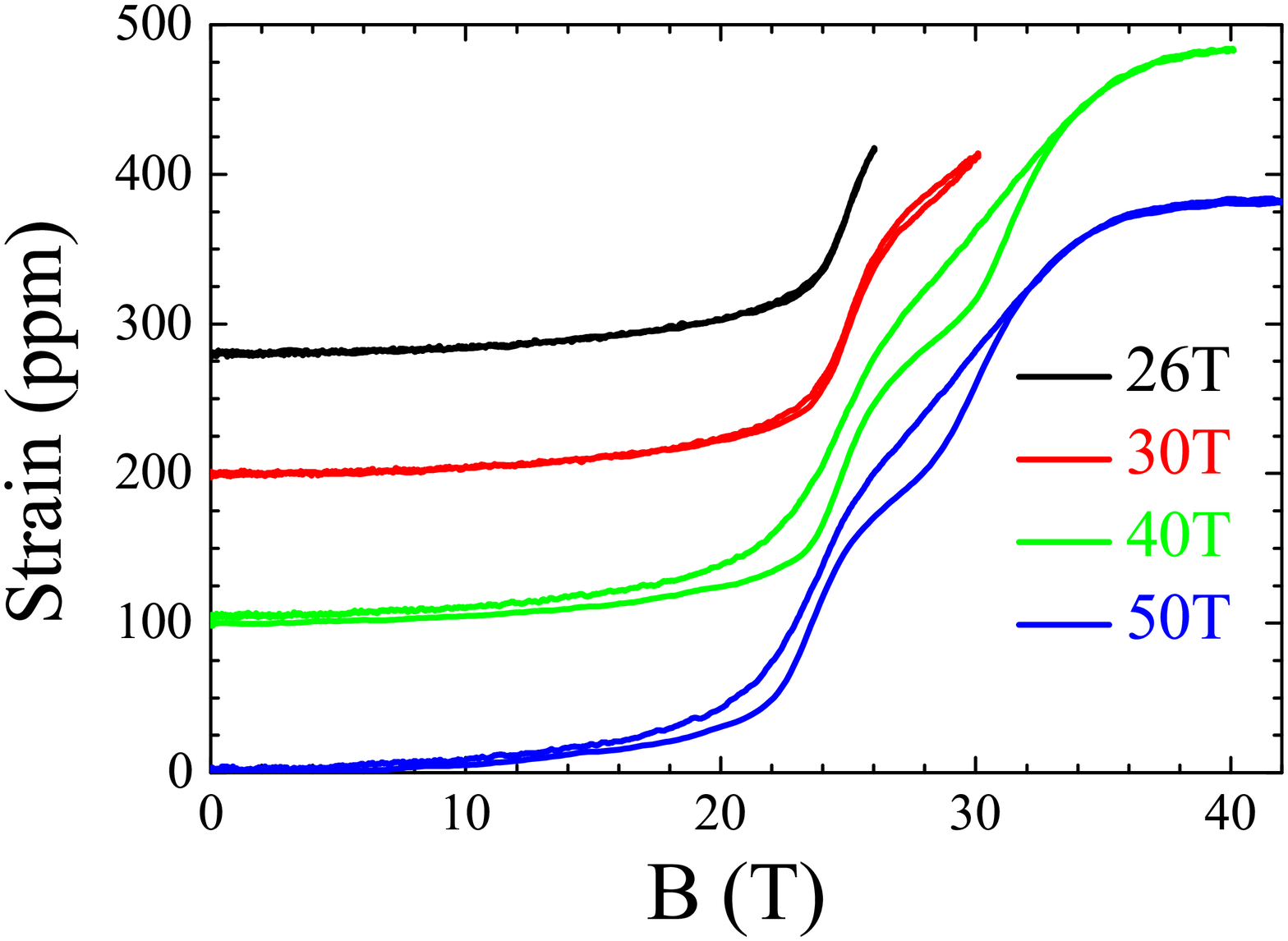}
\caption{(Color online) Axial magnetostriction of UPt$_2$Si$_2$ for magnetic fields $B \parallel c$ axis at 1.8\,K; data shifted for clarity. The plot illustrates the absence/appearance of structural hysteresis upon ramping the field into phase $III/V$, respectively. The legend denotes the highest magnetic fields attained for the different magnet runs; for details see text.}
\label{fig:MShys}
\end{figure}

Similar to the magnetoresistivity, we find a history dependent hysteresis at low $T$ (Fig.~\ref{fig:MShys}): For measurements from zero field up into phase $III$ and back no hysteresis is observed. Conversely, when the final field lies within phase $V$, structural hysteresis appears. Thus, the structural probe magnetostriction verifies that the phase transition $III - V$ is of a first order nature.

\begin{figure}[htb]
\centering
\includegraphics[width=1 \columnwidth]{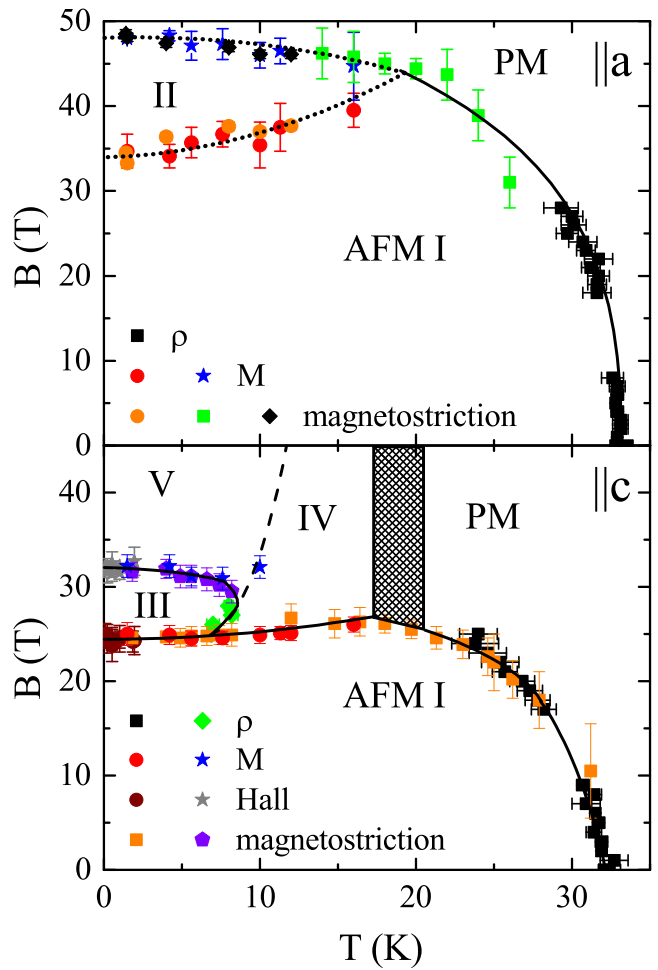}
\caption{(Color online) The magnetic phase diagrams of UPt$_2$Si$_2$ for the field $B \parallel a$ and $c$ axes from combining the present data with those reported in Ref.~[\onlinecite{schulze2}]; for details see text.}
\label{fig:pd}
\end{figure}

With the present data set we complete our high field studies on UPt$_2$Si$_2$. By combining the new data with those from Ref.~[\onlinecite{schulze2}], we present the magnetic phase diagrams of UPt$_2$Si$_2$ for $B \parallel a$ and $c$ axes in Fig.~\ref{fig:pd}. Altogether, our new set of data fully confirm the essential findings on the magnetic phase diagrams of UPt$_2$Si$_2$ as reported in Ref.~[\onlinecite{schulze2}]. In particular, for the field $B \parallel a$ axis, the new data points derived from magnetostriction measurements, which define the phase border lines from the AFM phase $I$ into the hysteretic regime $II$ and the paramagnetic regime sit well on top of those previously established. Furthermore, in the intermediate temperature regime $\sim$\,20\,--\,30\,K the new data now define the border lines more accurately as was possible with the data presented in Ref.~[\onlinecite{schulze2}]. 

For the field $B \parallel c$ axis at low temperatures the border lines between the phases $I$, $III$ and $V$ are perfectly reproduced with our new data. Furthermore, the existence and nature of the phase border lines has been established down into the mK-range. Finally, with the new data the phase border lines are now more accurately determined in the intermediate temperature regime $\sim$\,15\,--\,25\,K.

As we have noted before \cite{schulze2}, we believe that there must be additional phase border lines in high magnetic fields between the paramagnetic regime and a phase $IV$, and between phase $IV$ and $III/V$. These observations are based on the qualitative change of the field dependent character of the magnetization (see Fig. 1 with field $B \parallel c$ axis in Ref.~\cite{schulze2}). First, the magnetization at 20\,K and above evolves monotonically in a Brillouin-function like fashion with field, while for the data taken at 16\,K and below there is a hysteretic metamagnetic transition at around 25\,T. Therefore, the metamagnetic transition must be into a phase different from the paramagnetic regime, {\it viz.}, into phase $IV$. As well, the magnetization taken in the temperature range $\sim$\,10 to 16\,K exhibits a single metamagnetic transition, while for data at 10\,K and below there is a two-step transition (the same has been observed in the magnetostriction, see above). In a similar line of argumentation, it suggests a transition into a new phase $V$ at low temperatures.

From our previous data we could only roughly estimate the position of these phase border lines and the associated tricritical points. With our magnetostriction data we can more accurately define the evolution of the phase border lines, as is done in Fig.~\ref{fig:pd}. Notably, the new data suggest that for $B \parallel c$ axis the border line of phase $I$ exhibits a shallow maximum around $\sim$\,20\,K/25\,T. Such behavior would be highly unusual for a common antiferromagnet. Instead, it appears that the competition with the high field phase $IV$ produces this anomalous evolution of the phase border line. Hence, our new data are fully consistent with the phase diagram scenario labelled ''A'' in Ref.~\cite{schulze2}. Within this scenario, we conclude that the upper tricritical point lies at around 25\,T and 19\,K. Unfortunately, close inspection of our various data sets utilizing different experimental tools does not provide a clear-cut signature unambiguously defining the tricritical point. As well, the precise evolution of the border between phase $IV$ and $V$ is rather awkward, as is the detailed structure of the area around the tricritical phase $III-IV-V$ point. Ultimately, to unambiguously  establish and define these phase border lines, and to definitely discard scenario ''B'' from Ref.~\cite{schulze2}, it would still require experiments in high magnetic fields carried out as function of temperature. Unfortunately, given the high field range, such experiments are rarely carried out.

\section{Band structure calculations}

For $B \parallel c$ axis the experimental evidence is consistent with the transition into phase $V$ to be an ETT in a correlated electron system. We have observed a significant change of the Hall coefficient at the phase $III-V$ transition for $B \parallel c$ axis, in line with a Lifshitz type character. As well, the first order nature of this transition is consistent with a Lifshitz type transition for an interacting electron system. Now, as a next step, and to complement our experimental study, we have carried out additional band structure calculations, aiming to identify features in the band structure beyond those established in Ref.~[\onlinecite{Elgazzar12}].

For an ETT, the topological changes in the iso-energy surfaces result from critical points in the band dispersion, {\it i.e.}, minima, saddle points, and maxima which give rise to van Hove singularities in the density of states. The changes in the topology of the iso-energy surfaces include the appearance or disappearance of small pockets, the formation of voids and the disruption of necks. Therefore, the focus of the present calculations is on critical points in the quasiparticle dispersion of UPt$_{2}$Si$_{2}$. For magnetic field-induced Lifshitz transitions, the critical points have to be rather close to the Fermi energy. All in all, it is thus the occurrence of these pockets, voids or necks that we are searching for in the band structure.

The present calculations assume that there are itinerant $5f$ electrons which form partially filled coherent bands. We analyse the Fermi surface where the energy bands are calculated under the following assumptions about the nature of the $5f$ electrons: We begin by adopting DFT treating all $5f$ electrons as band states. This approximation scheme cannot fully capture the correlation effects. To simulate the latter we calculate the band structure under the assumption that two of the $5f$ electrons are localized while one may be itinerant and hybridize with the conduction states. For simplicity, we first treat all $5f$ channels as equivalent and account for orbital selection in a second step. For the last step, we single out the $5f$ electron in the $j=5/2$, $j_{z}=\pm1/2$-channel as the hybridized electron. 

More specifically, the band structure results reported in the present paper were obtained by the fully relativistic formulation of the linear muffin-tin orbitals (LMTO) method \cite{Andersen75,SkriverBook,Christensen84,Albers86}. The spin-orbit interaction is fully taken into account by solving the Dirac equation. The results are compared to the relativistic calculations by Elgazzar et al. \cite{Elgazzar12}. Because the heavy fermion compound UPt$_{2}$Si$_{2}$ crystallizes in the tetragonal CaBe$_{2}$Ge$_{2}$ structure, the crystal structure is relatively open and, consequently, the atomic-sphere approximation (ASA) cannot be expected to give a sufficiently accurate description of the electronic band structure. The combined correction term which contains the leading corrections to the ASA alters \cite{Andersen75} the conduction bands in a characteristic way and hence cannot be neglected. Exchange and correlation effects were introduced using the Barth-Hedin potential \cite{Barth72}. The band structure was converged for 405 ${\bf k}$-points in the irreducible wedge, whose volume equals 1/16 of the Brillouin zone. The density of states (DOS) was evaluated by the tetrahedron method with linear interpolation for the energies \cite{Jepsen71,Lehmann72}. For the conduction band the DOS was calculated at 0.25 mRy ($\approx$ 0.0034 eV) intervals.

The effective potential seen by the conduction states is approximately constructed as a superposition of contributions, which have spherical symmetry inside atomic and ''empty'' spheres surrounding lattice or interstitial sites, respectively. The empty spheres should be viewed as auxiliary constructions that permit an improved description of the electron density as well as the potential within the framework of the ASA. In UPt$_{2}$Si$_{2}$  the dominant contribution to the charge in the interstitial region comes from the Pt-$d$ states. 

Our calculations are done at the experimental lattice parameters and do not correspond to the equilibrium geometry of an LDA calculation. The total energy evaluated for the experimental structure will therefore exceed its theoretical minimum value. This difficulty, which is generally encountered in metals with strongly correlated electrons, is a direct consequence of the LDA description of these systems in terms of a single-particle picture. This can be seen by considering two limiting cases. First, treating the $f$ electrons as part of the ion core implies that their contribution to binding is neglected. As a consequence, the equilibrium values of the lattice constants are often overestimated. Secondly, describing the $f$ electrons as band electrons yields a relatively narrow, partially filled $f$ band at the Fermi level. The calculated LDA DOS at the Fermi level is large compared with that of ordinary metals. An effective single-particle description such as the LDA predicts an electronic compressibility which is enhanced over that of ordinary metals by the same factor. The behavior anticipated for independent fermions, however, is in contradiction with experiment which yields compressibility values for heavy fermion metals which are comparable to those of ordinary metals. The large electronic compressibility predicted erroneously by the LDA in $f$-metals leads to overbinding ({\it i.e.}, the theoretical values of the equilibrium lattice constants are too small). It is a direct consequence of the failure of the independent particle picture.

The calculations were done using two energy panels; {\it i.e.}, two separate LMTO calculations were performed to determine self-consistently the uranium 6$p$ states and the conduction bands, respectively. Treating the U-6$p$ semi-core states as band states accounts for the small overlap between these core states. The resulting narrow bands far below the Fermi energy only hybridize weakly with the conduction bands. This hybridization is then neglected in our method, but its influence on the shape of the potential is taken into account. The charge contributions of the other core states were taken from atomic calculations and kept frozen during the iterative procedure. For the lower panel we included $s-p-d$ angular momentum components in the basis at the U and Pt sites and $s-p$ components at the Si and interstitial sites. For the upper panel we included $s-p-d-f$ components at the U site and $s-p-d$ on the remaining sites.

\begin{figure*}
\includegraphics[width=0.25\textwidth]{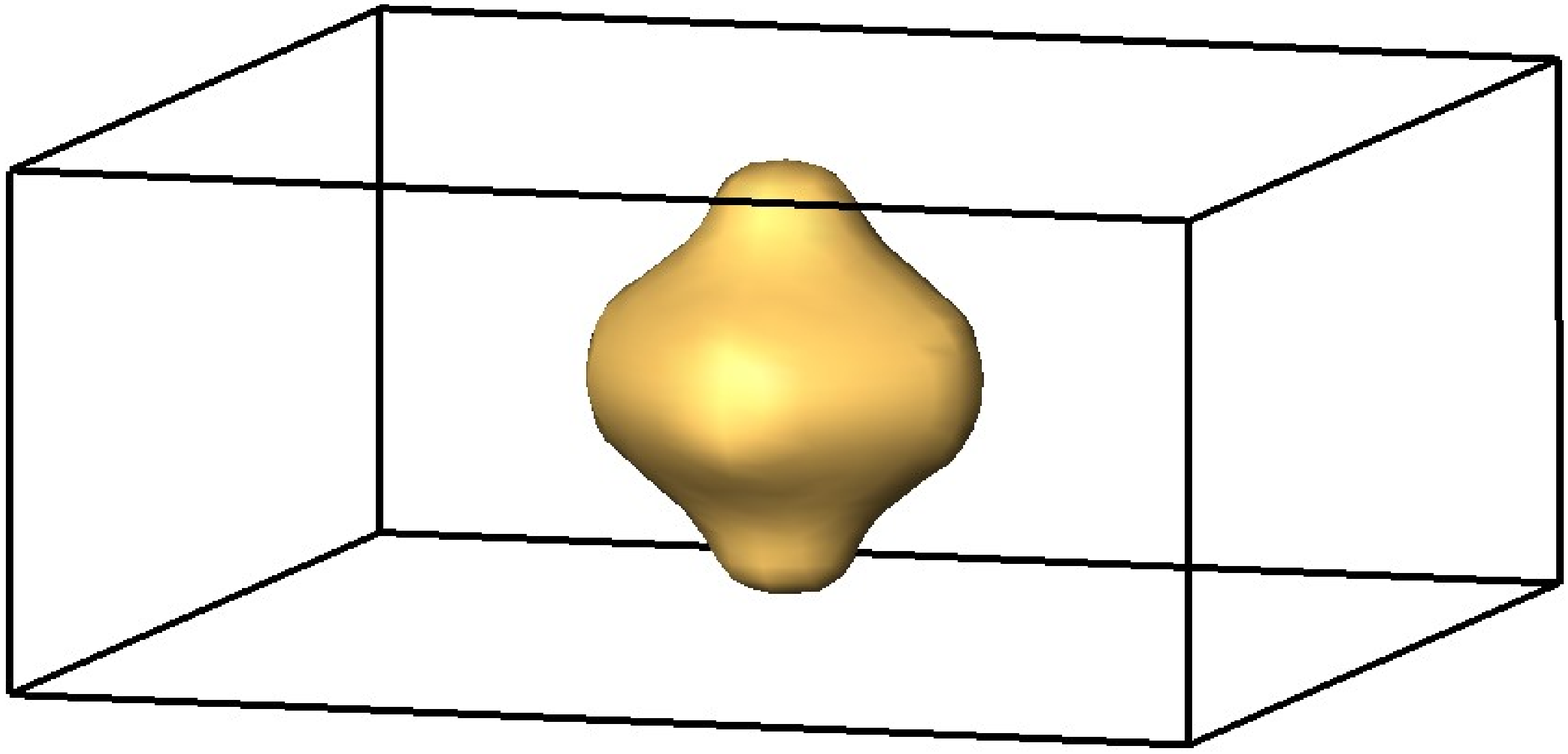}\includegraphics[width=0.25\textwidth]{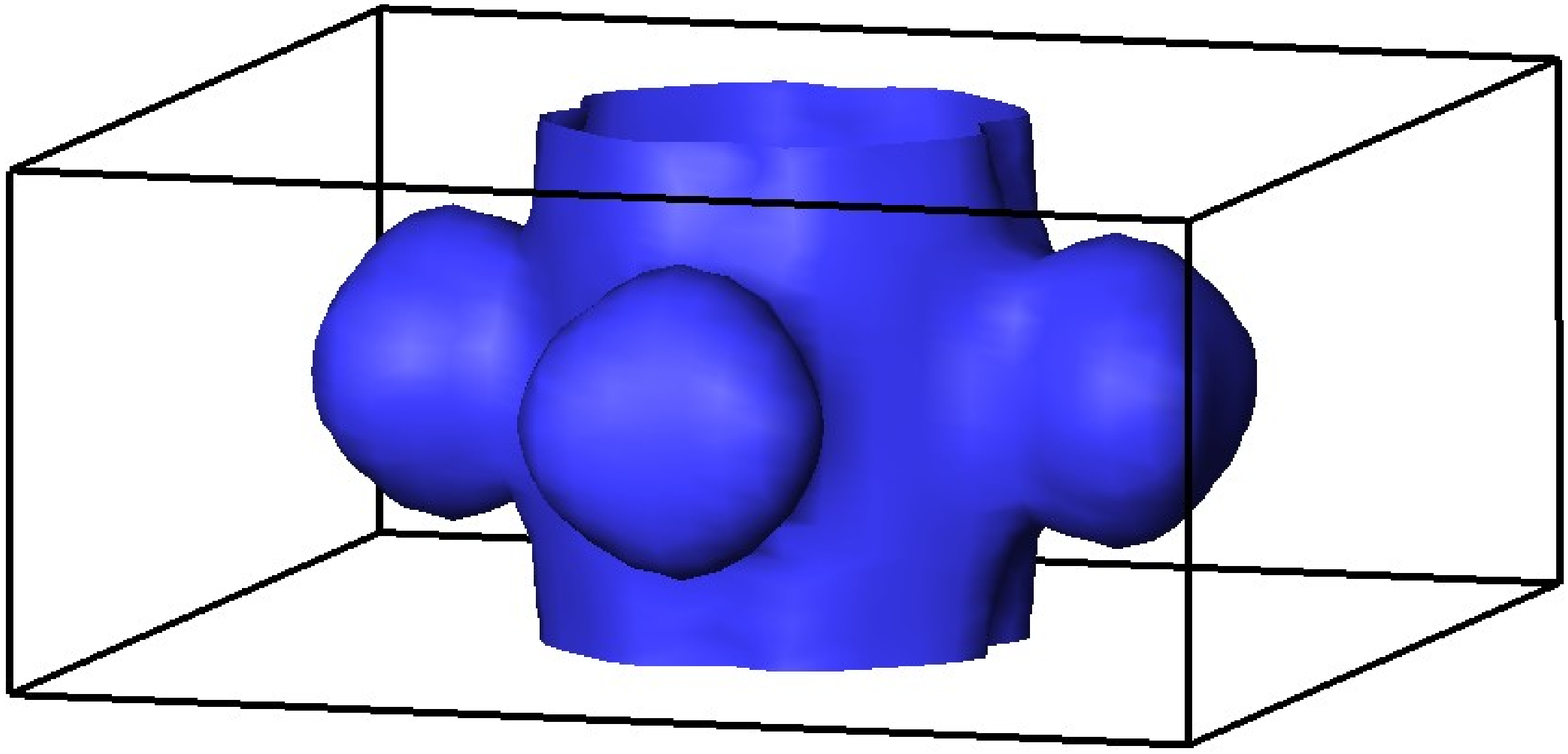}\includegraphics[width=0.25\textwidth]{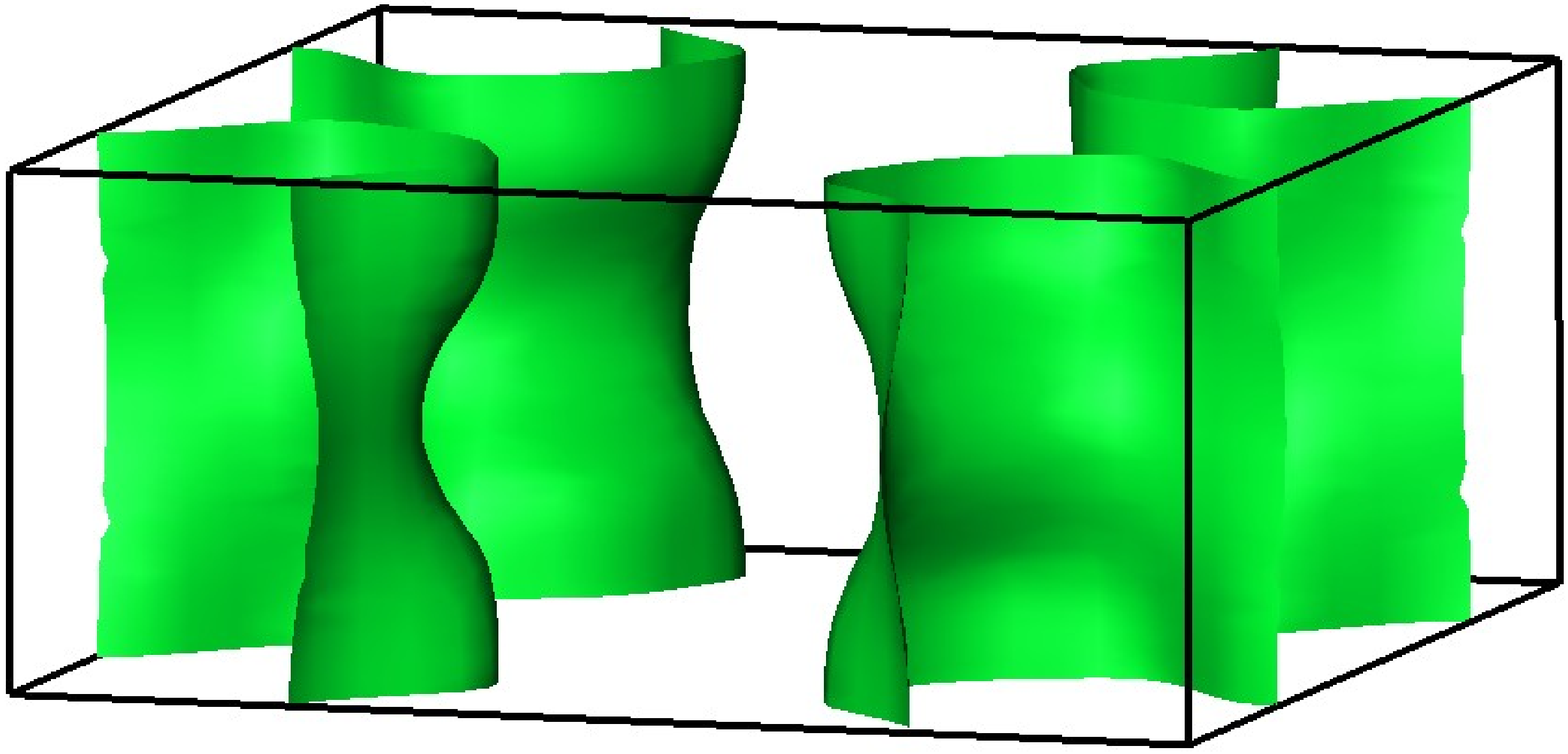}\includegraphics[width=0.25\textwidth]{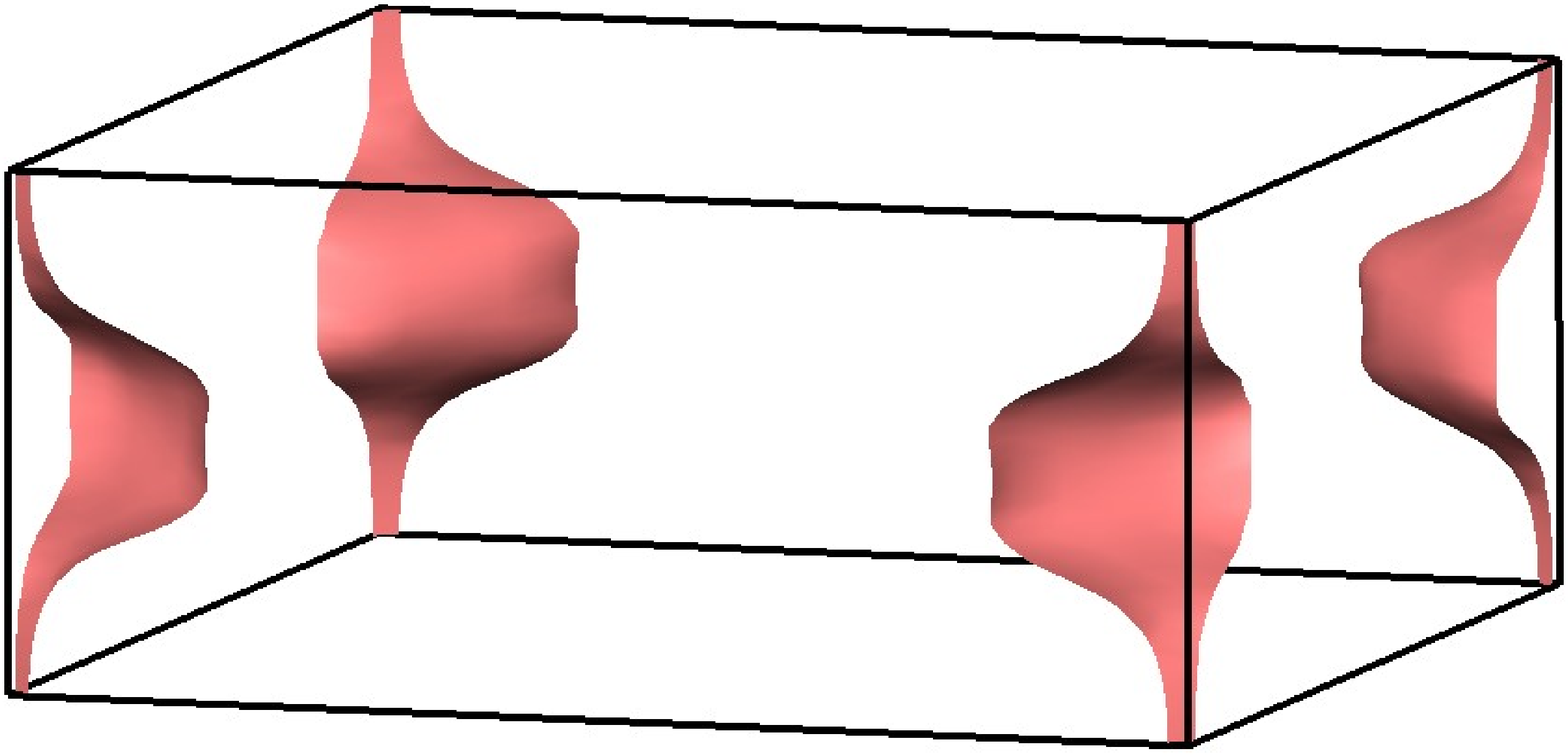} \\
\includegraphics[width=0.25\textwidth]{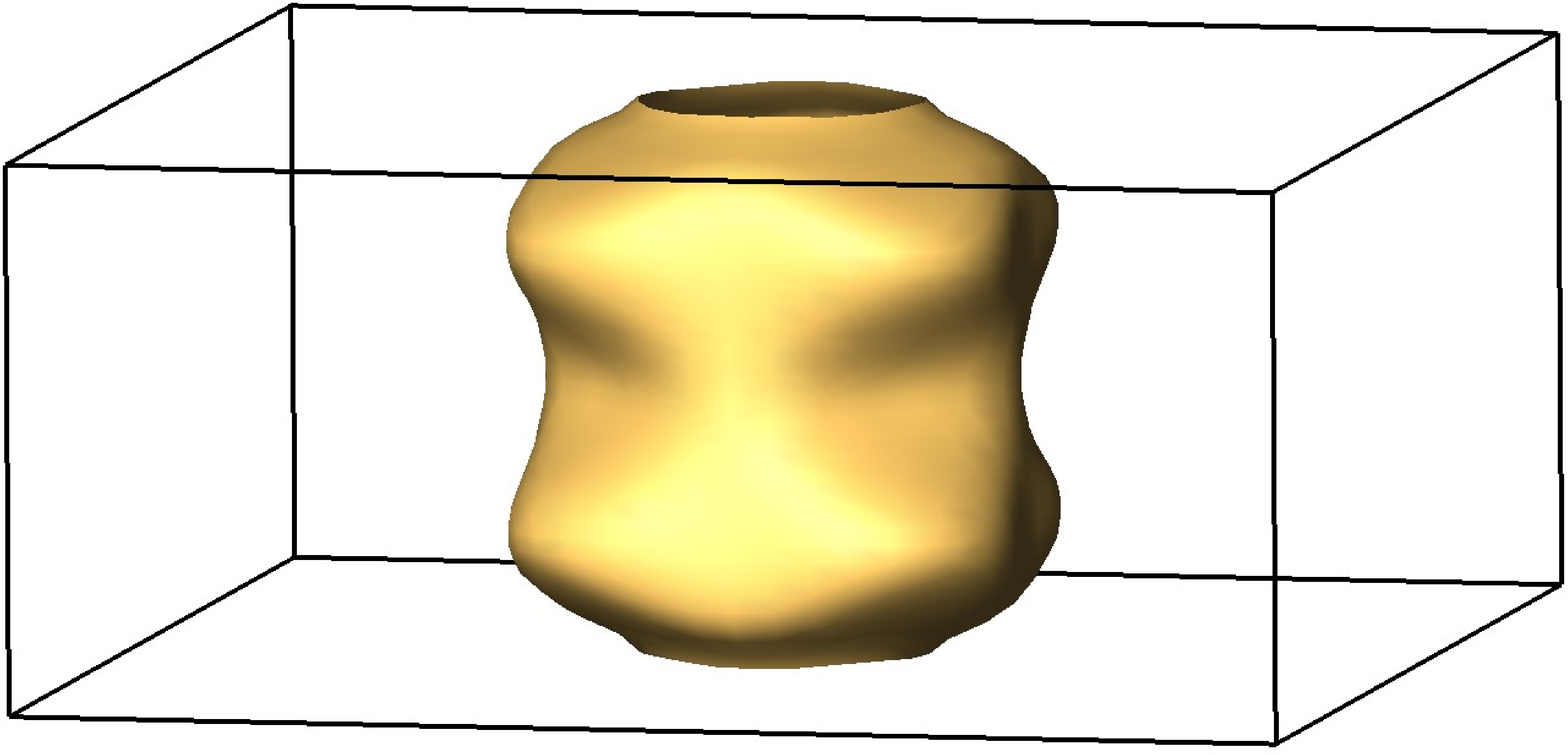}\includegraphics[width=0.25\textwidth]{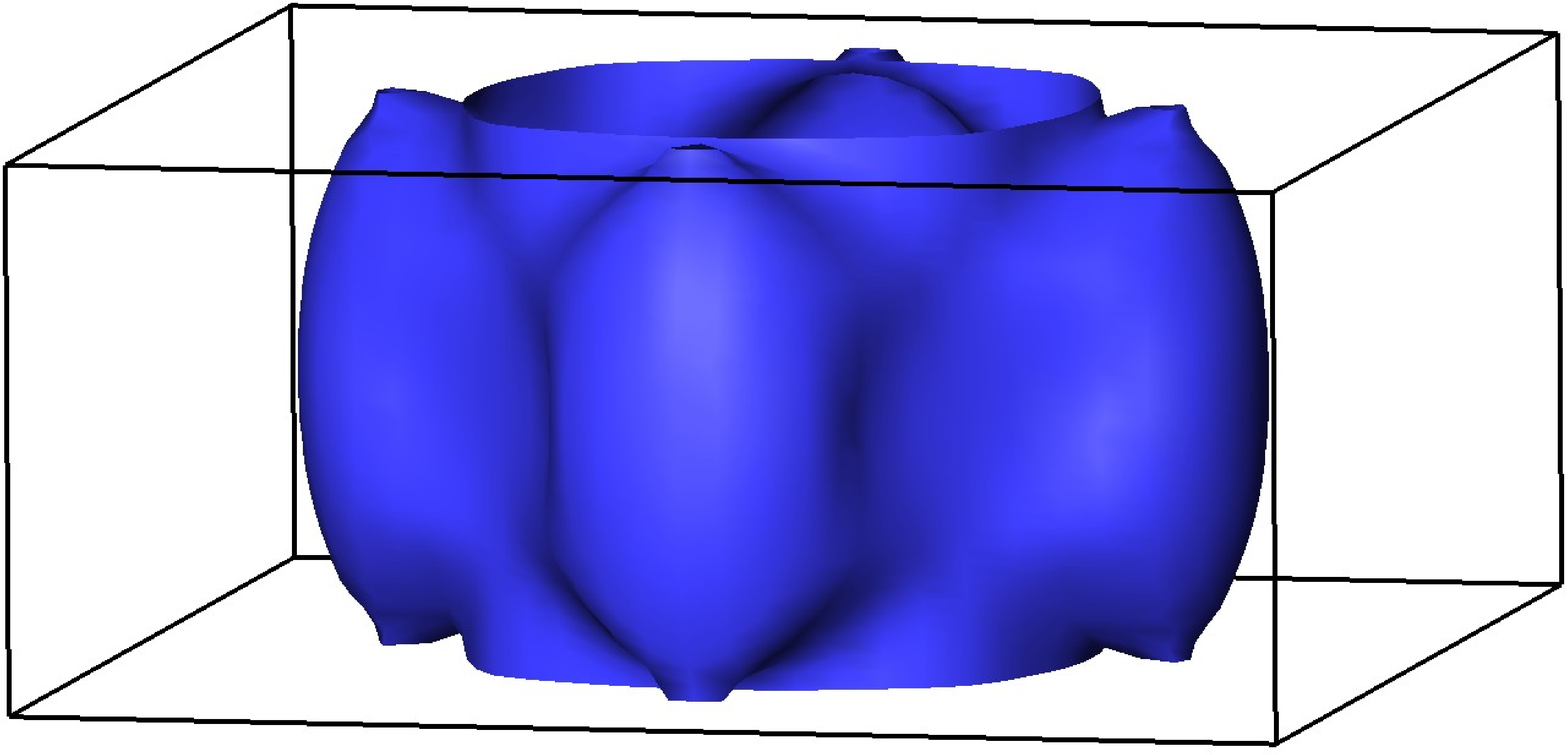}\includegraphics[width=0.25\textwidth]{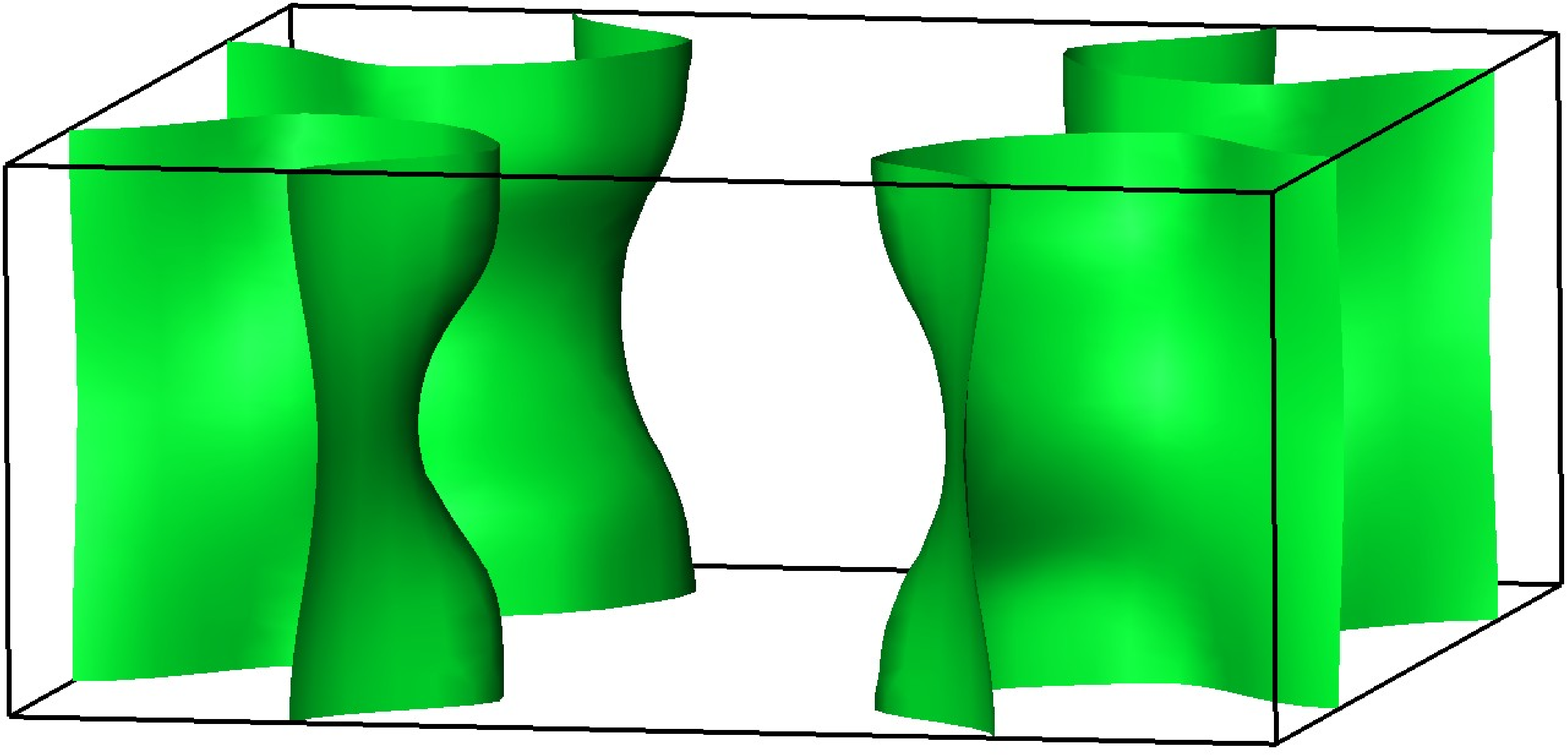}\includegraphics[width=0.25\textwidth]{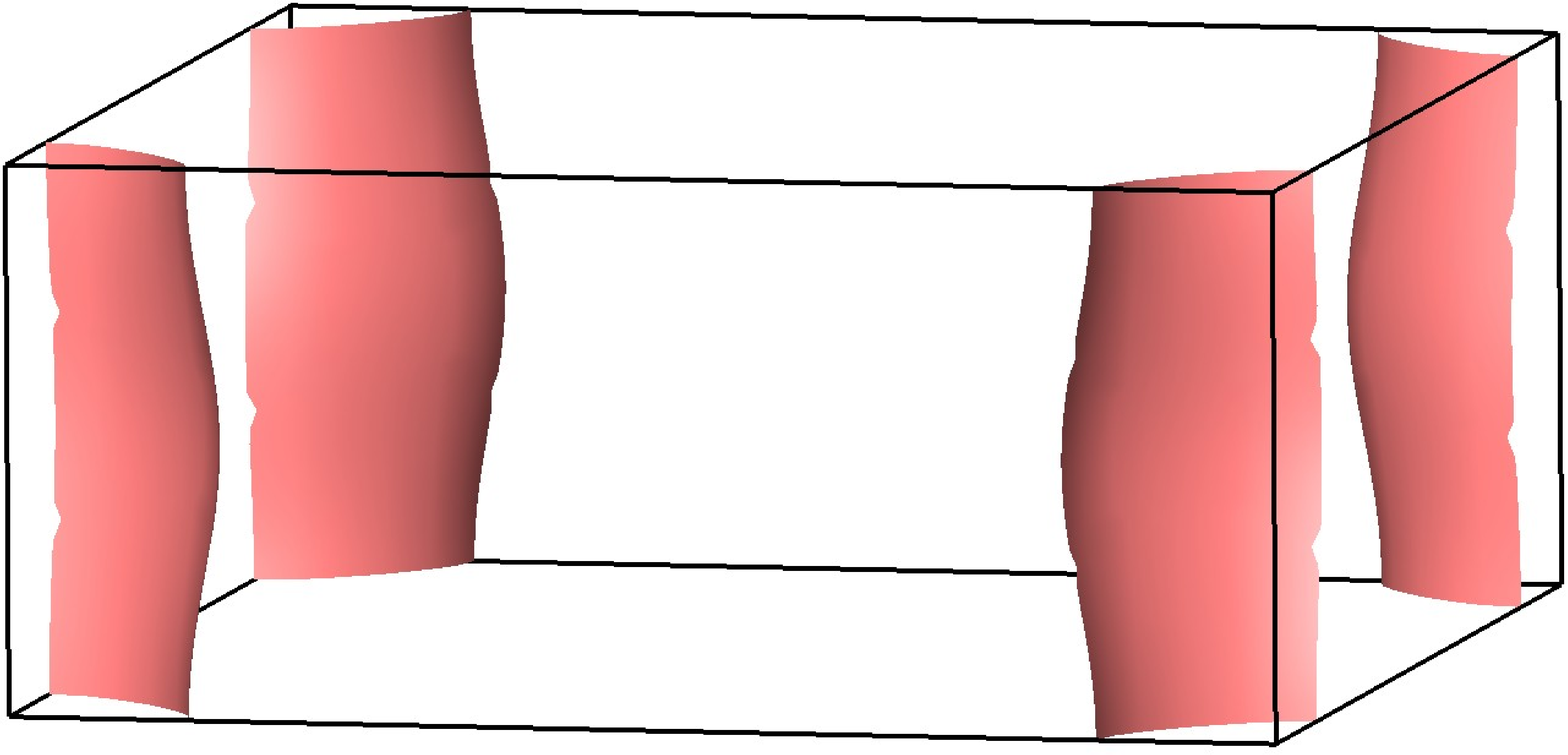} \\
\includegraphics[width=0.25\textwidth]{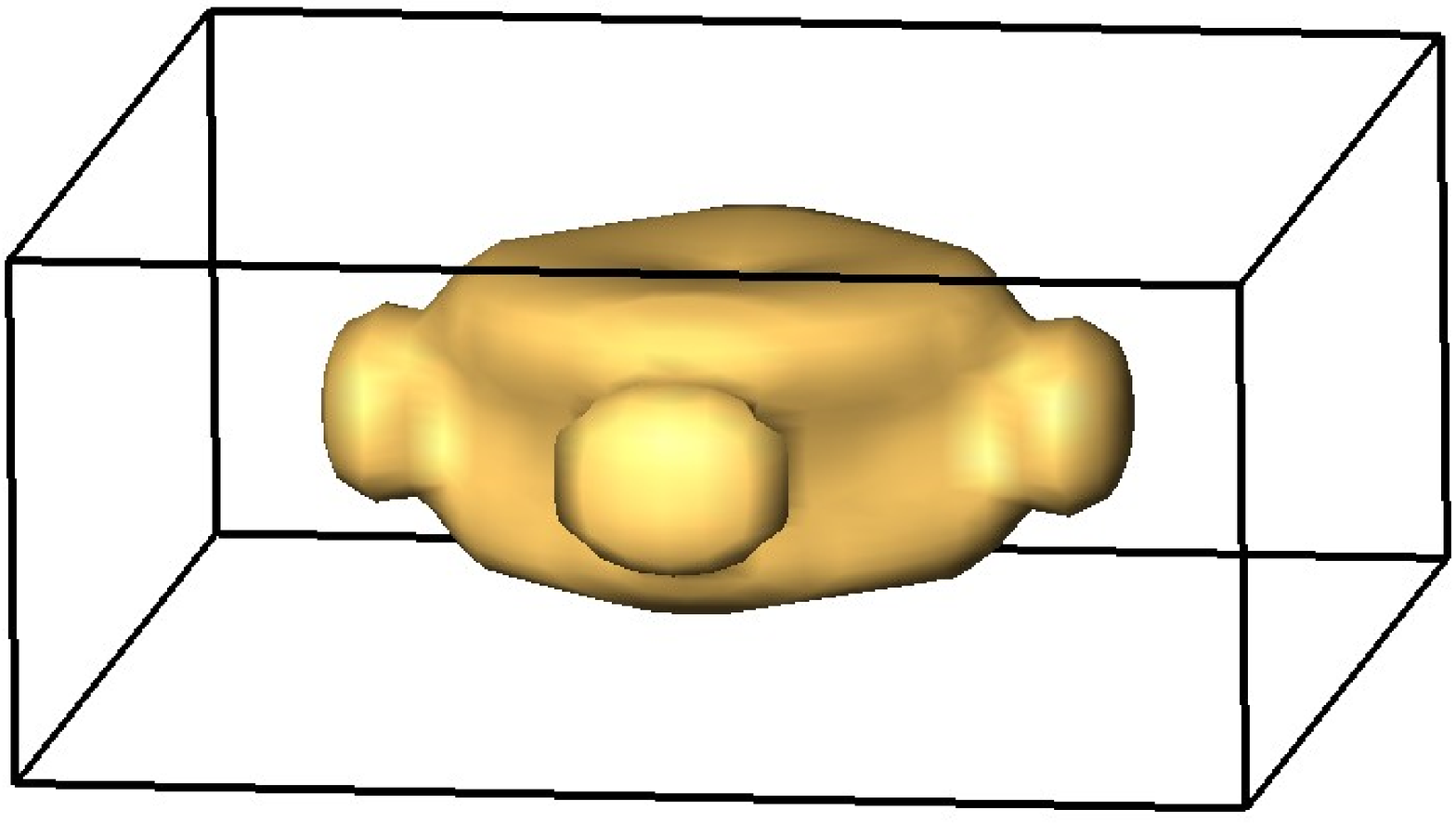}\includegraphics[width=0.25\textwidth]{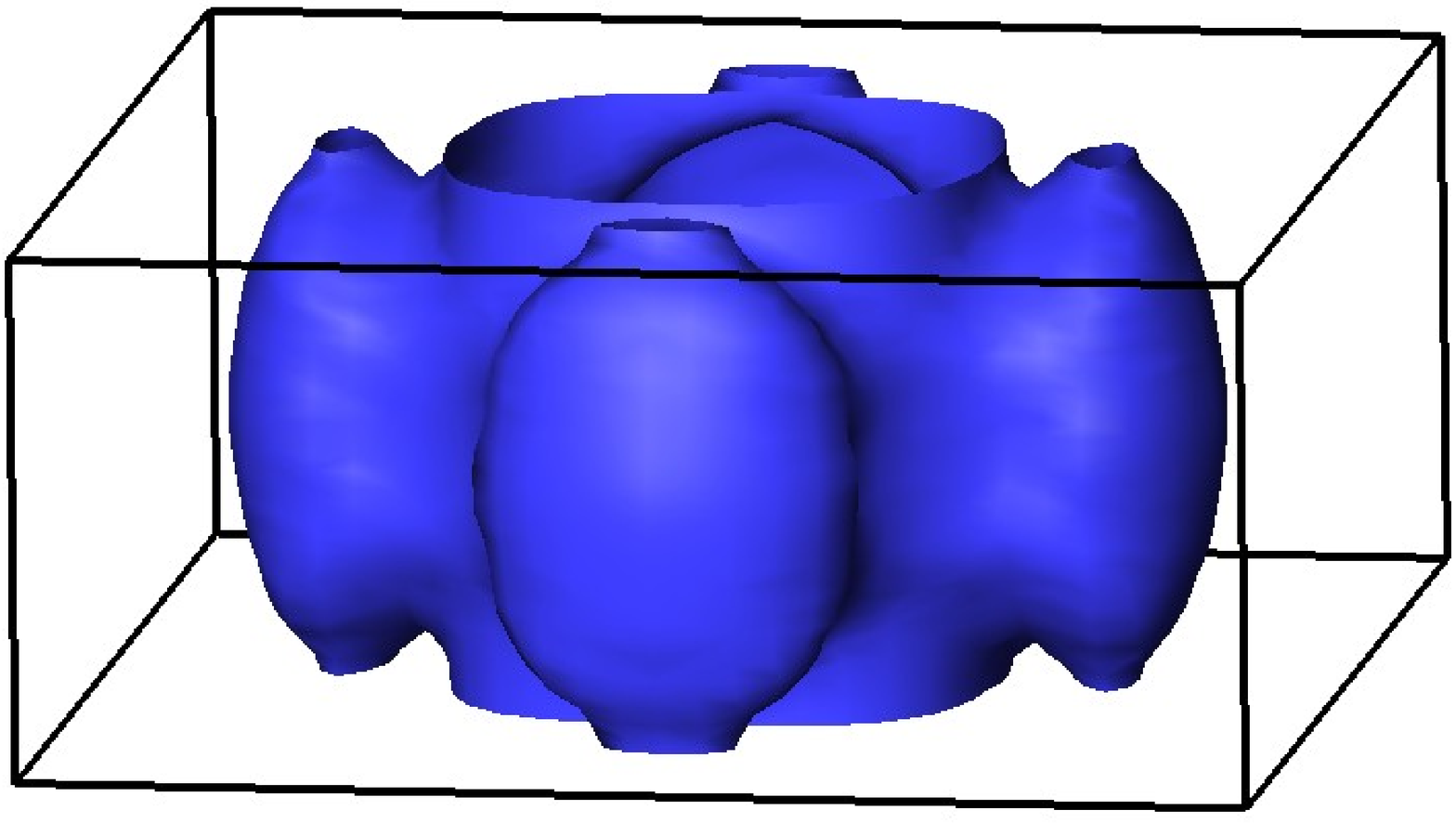}\includegraphics[width=0.25\textwidth]{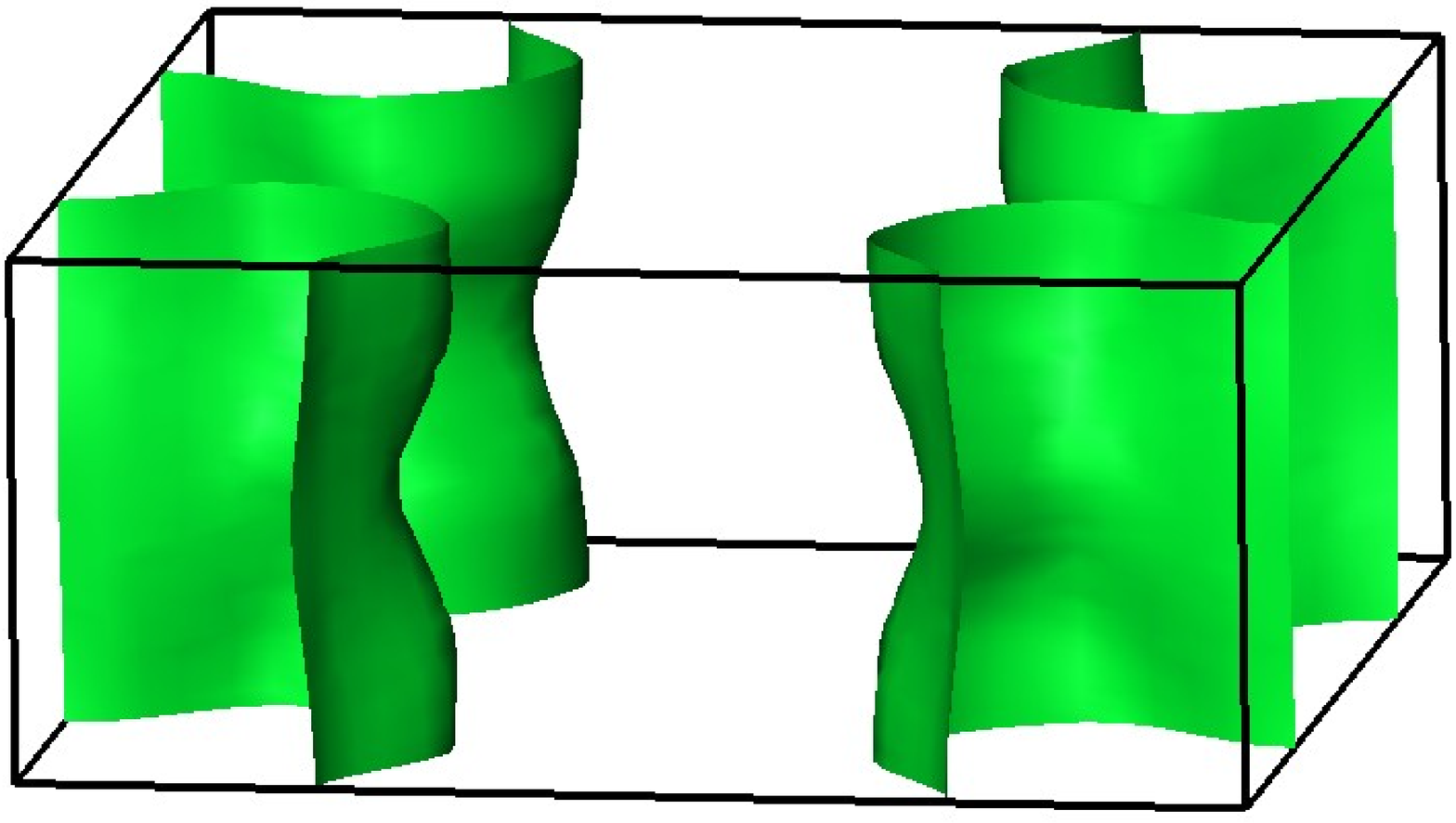}\includegraphics[width=0.25\textwidth]{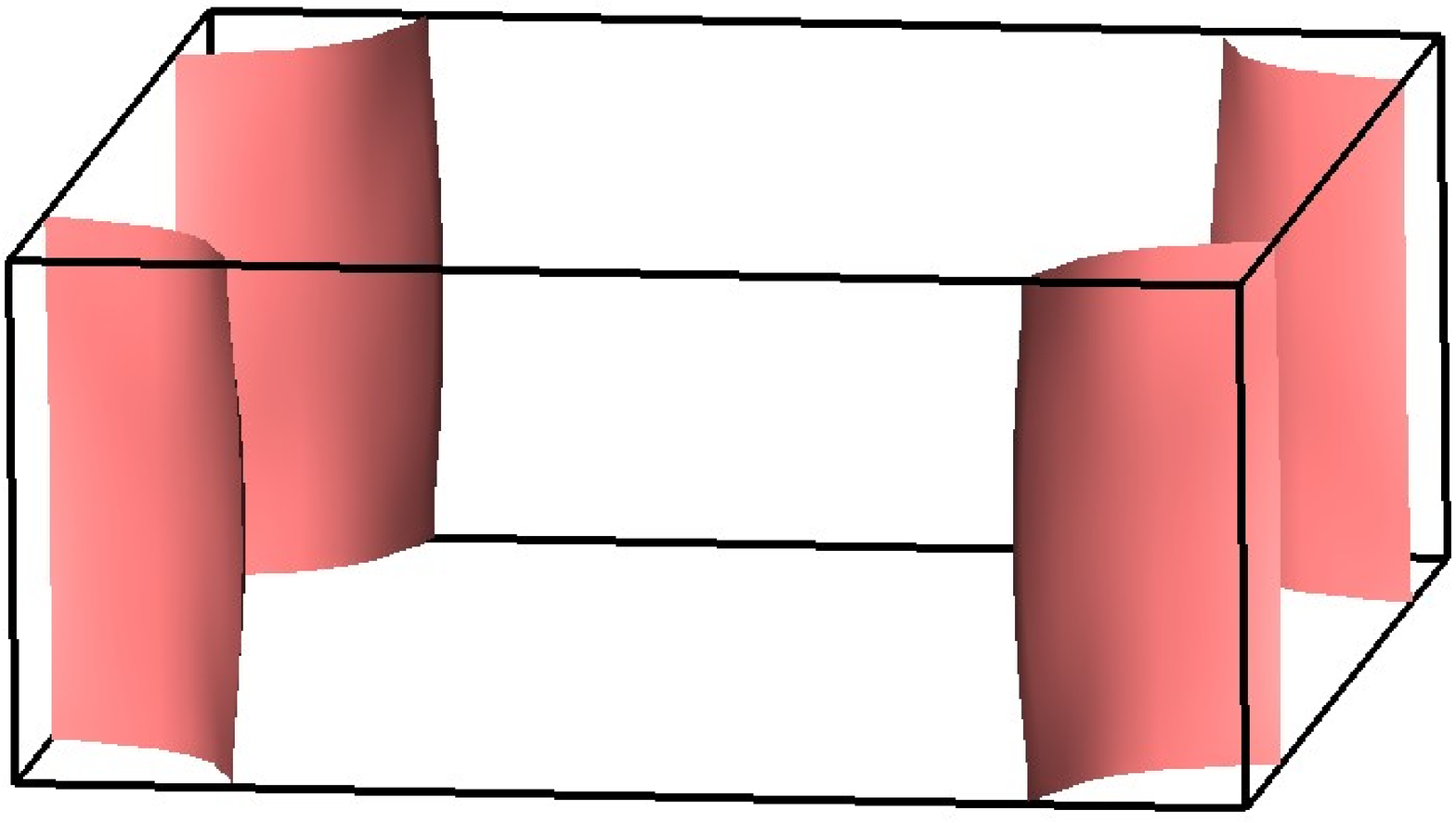}
\caption{Calculated Fermi surfaces of UPt$_{2}$Si$_{2}$ in the paramagnetic phase. Top row: LDA calculation treating all $5f$ electrons as itinerant band electrons. Second row: Two $5f$ electrons are treated as part of the ion core while one $5f$ electron hybridizes with the conduction electrons. Third row: Influence of orbital-selective localization is accounted for by treating two $5f$ electrons as part of the ion core while one $5f$ electron in the $j=5/2$, $j_{z}=\pm1/2$-channel hybridizes with the conduction electrons.
\label{fig:UPt2Si2FermiSurfaces}}
\end{figure*}

The Fermi surface obtained by the three approaches regarding the degree of localization of the 5$f$-electrons are summarized in Fig.~\ref{fig:UPt2Si2FermiSurfaces}. We find four bands crossing the Fermi energy, and denote the corresponding FS sheets as 1, 2, 3 and 4. Globally, the LDA result agrees well with the one obtained by Elgazzar et al. \cite{Elgazzar12} apart from the fact that the ''appendices'' are absent in our sheet 1 (see Elgazzar band labelled 113). 

With respect to the FS topology, we find that the FS sheets 3 and 4 are remarkably insensitive to the treatment of the $5f$ states. The number of itinerant $5f$ electrons affects only the sheets 1 and 2. Correspondingly, we have inspected closely the response of these sheets on magnetic field by analyzing the iso-energy surfaces for shifts away from the Fermi energy by $6$\,meV. For a magnetic moment $\sim 2.5$\,$\mu_B$ as in UPt$_2$Si$_2$ this value corresponds to a magnetic field of $\sim 30$\,T. 

\begin{figure*}
\includegraphics[width=0.33 \textwidth]{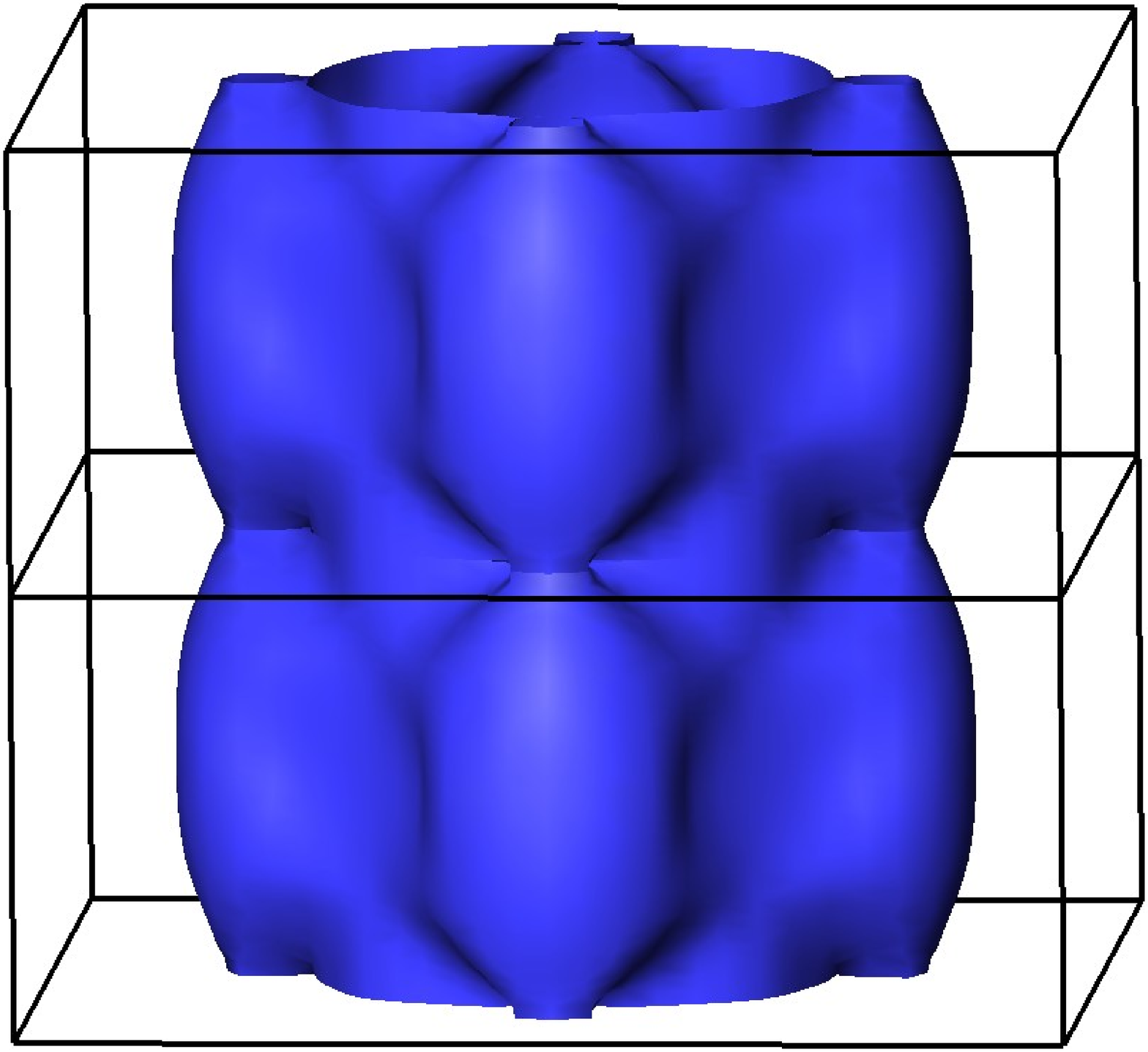}\includegraphics[width=0.33 \textwidth]{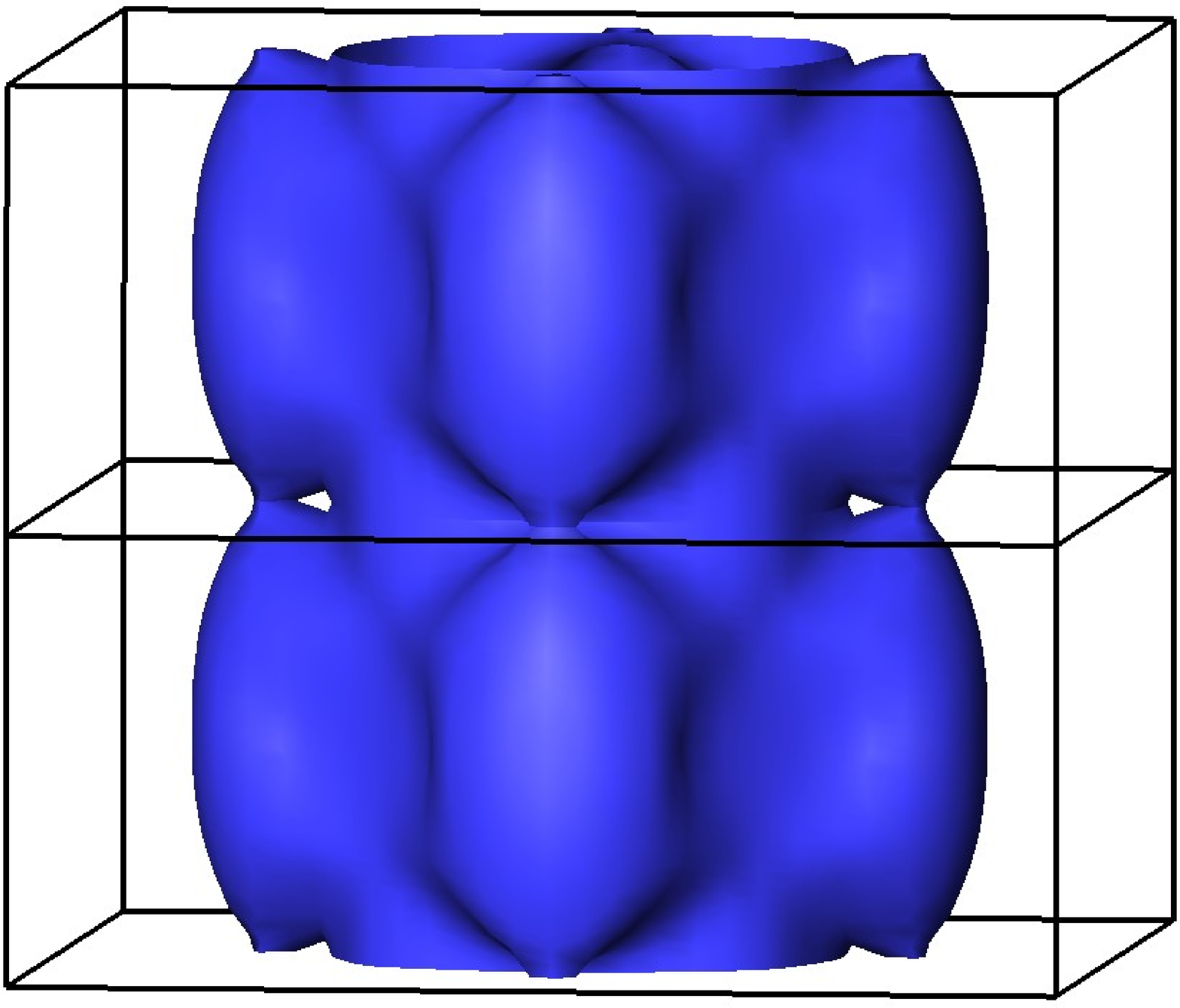}\includegraphics[width=0.33 \textwidth]{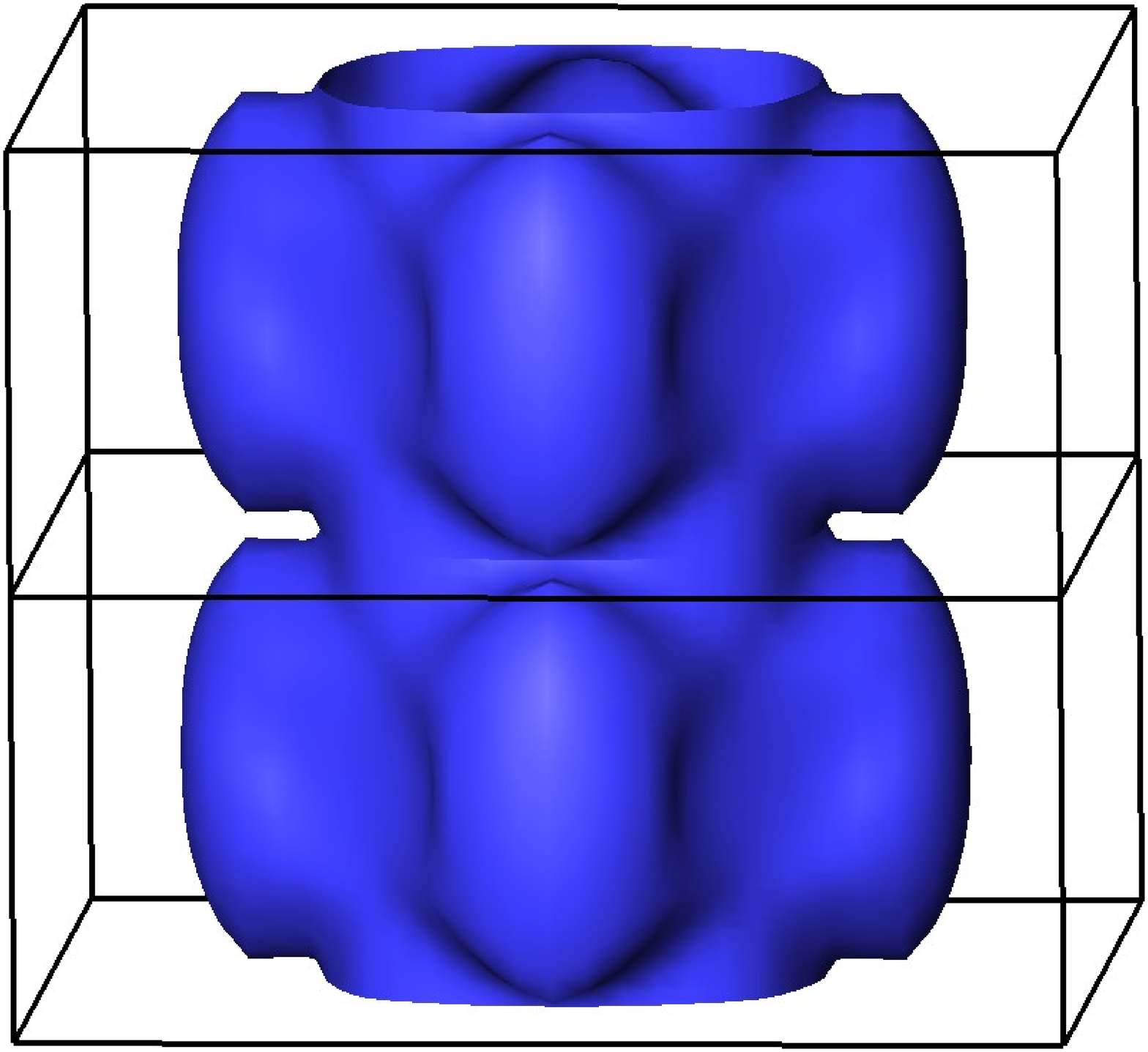}
\caption{Lifshitz transitions for the case of two $5f$ electrons treated as localized. The iso-energy surfaces $E=E_{F}-6$\,meV (left panel), $E=E_{F}$ (middle) and $E=E_{F}+6$\,meV (right) show a cascade of ETT by void formation and neck disruption, and which are accessible by magnetic fields in the range of a few 10\,T.}
\label{fig:LifshitzFSf2CoreSheet2}
\end{figure*}

In particular, for sheet 2 we find a qualitative change of the shape of the FS for such a small energy shift, thus providing direct band structure evidence for an ETT (see Fig.~\ref{fig:LifshitzFSf2CoreSheet2}). Clearly, a void formation/neck disruption is visible as the iso-energy surface is tuned from -$6$\,meV to +$6$\,meV around the Fermi energy. For FS sheet 1 in the energy range considered we find no topological change.

\begin{figure}
\includegraphics[width=0.5\linewidth]{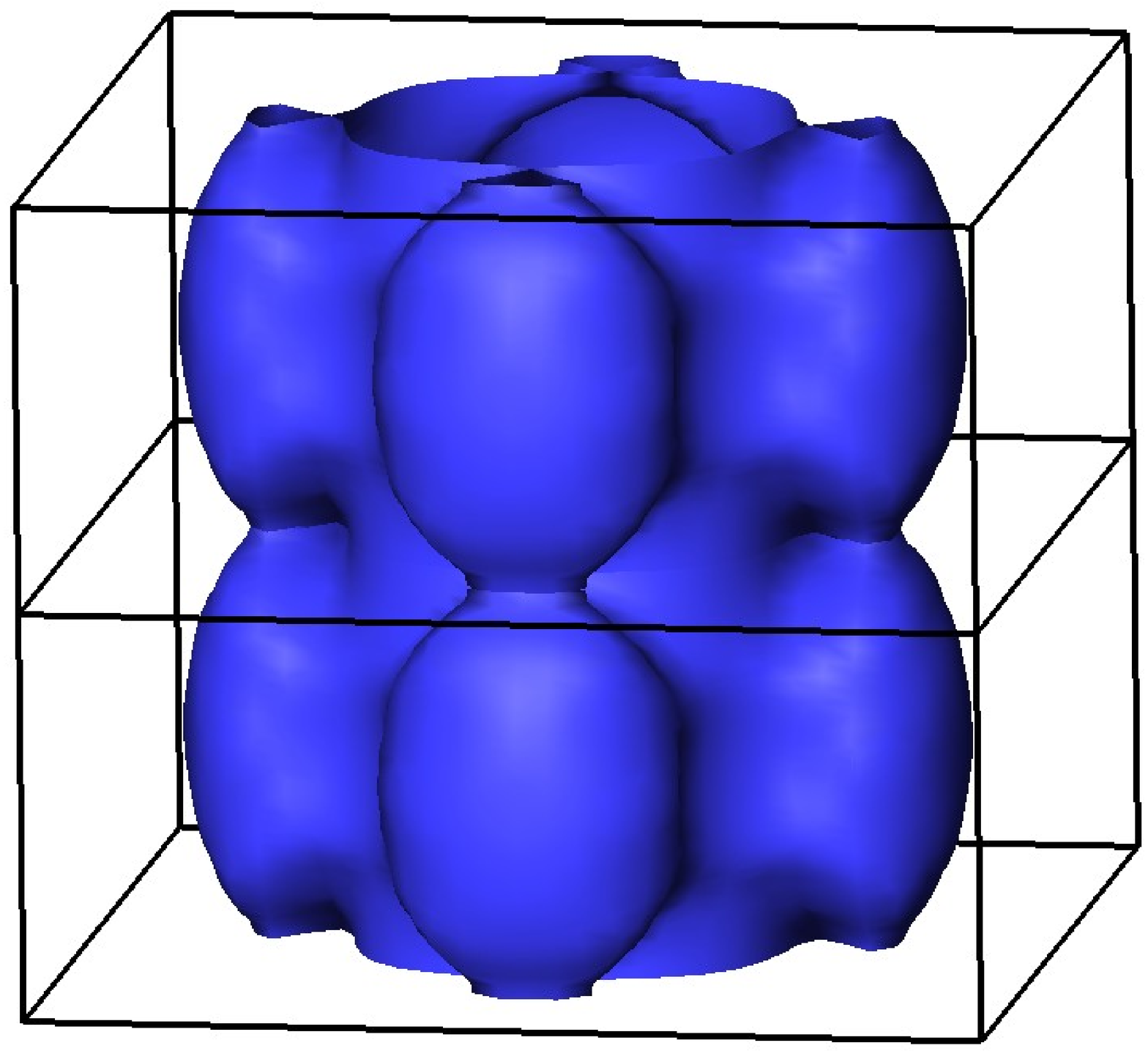}\includegraphics[width=0.5\linewidth]{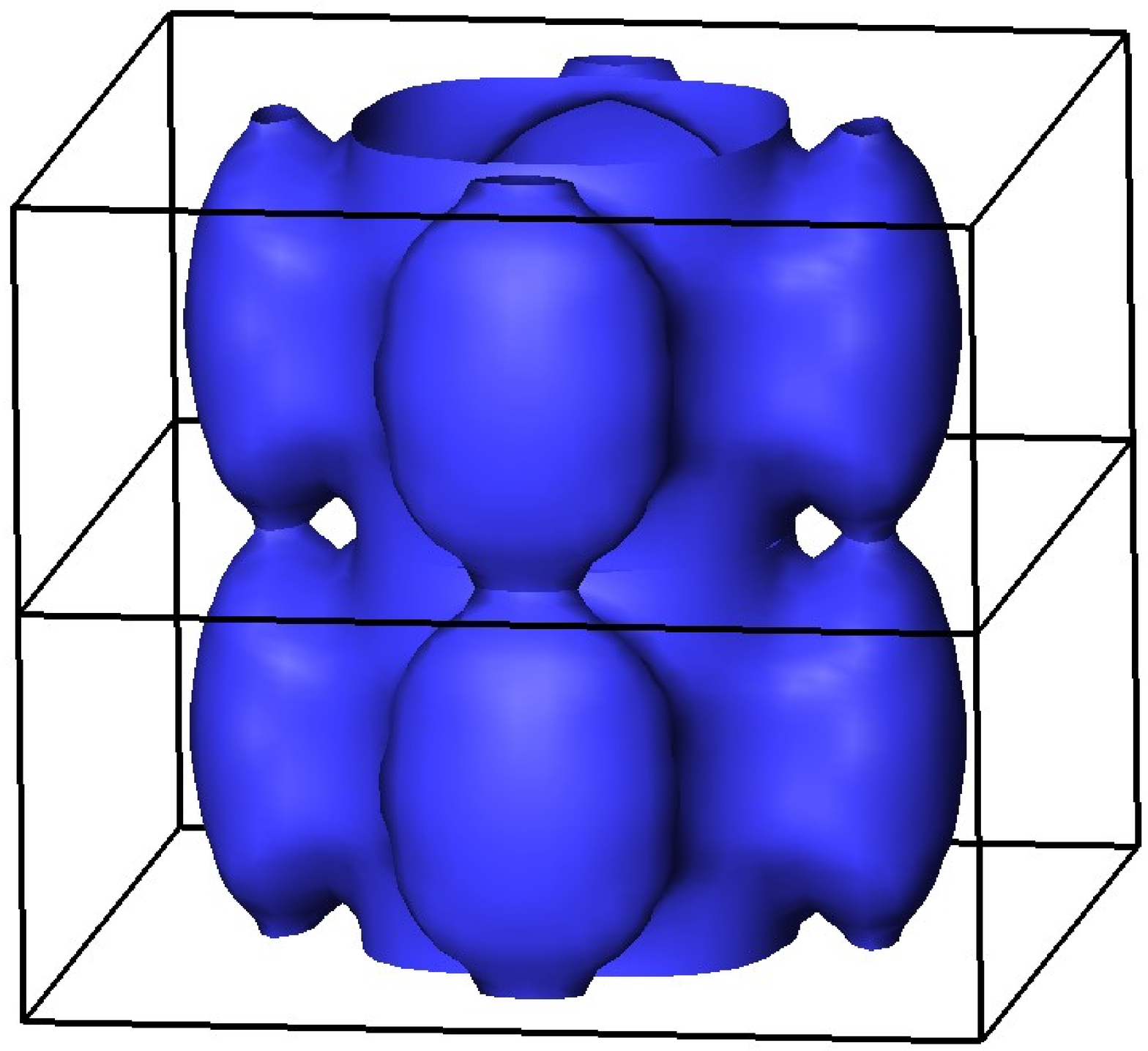}
\caption{ETT in the dual model treating two $5f$ electrons as localized and allowing the $5f$ $j=5/2$, $j_{z}=\pm1/2$ channel to hybridize with the conduction bands: Depicted are the iso-energy surfaces $E=E_{F}-6$\,meV (left panel) and $E=E_{F}+6$\,meV (right panel).}
\label{fig:LifshitzFSDualBand2}
\end{figure}

Following the identification of Fermi surface sheet 2 as the one being topologically affected by magnetic fields of the order of 30\,T, in a final step we have specified the character of the itinerant electron by allowing the $5f$, $j=5/2$, $j_{z}=\pm1/2$ channel to hybridize with the conduction bands. Again, the iso-energy surfaces shifted by $\pm 6$\,meV against the Fermi energy clearly reflect an ETT, as demonstrated in Fig.~\ref{fig:LifshitzFSDualBand2}. Thus, in our band structure calculations, and assuming one out of three $5f$ electrons being delocalized we find ETT on the Fermi surfaces of a correlated electron system, {\it viz.}, UPt$_2$Si$_2$, consistent with our experiments.  

\section{Conclusion}

In summary, we provide experimental evidence for the possibility of a field-induced first-order Lifshitz type transition in the correlated electron system UPt$_2$Si$_2$ through a combined study of the electronic and structural properties. Furthermore, for the FS, critical points close to the Fermi energy $E_{F}$ are found in the band dispersion when two of the $5f$ electrons are treated as localized, implying that field-induced Lifshitz transitions are to be expected. In contrast, for the all-itinerant model, the critical points leading to Lifshitz transitions are too far from the Fermi energy as to be relevant in an experimental context. Thus, with our study we demonstrate consistence of our experiments with the predictions made based on the dual model of $5f$ electrons for the case of an uranium intermetallic with strong electronic correlations.

Finally, the question arises about the nature of the other magnetic phases in UPt$_2$Si$_2$ for $B \parallel c$ axis. When associating the first-order transition into phase $V$ with an ETT, oppositely the second-order character of the phase $I - III$  transition would signal a more ordinary type of transition. The character of the transition, as seen in the magnetization (see Ref.~\cite{schulze2}), together with the size of the jump of the magnetization, could be consistent with for instance a spin-flop transition. In turn, this observation raises questions about the character of phase $IV$, as it shows up in the magnetization in a similar fashion as the $I - III$ transition. In terms of the magnetization, the difference between phases $III$ and $IV$ is not obvious.

Conversely, following a different line of arguments, and while thermal smearing might prohibit a definite identification, conceptually, phase $IV$ can have the FS topology of phase $I/III$, $V$ or a different one. If the FS topology is not that of phases $I/III$, consequently, there would be multiple Lifshitz transitions in the phase diagram for $B \parallel c$ axis of UPt$_2$Si$_2$. Taking this observation into a more general context, the interplay of spin reorientation/anisotropy and FS topology may give rise to a complex set of field-induced phases in UPt$_2$Si$_2$, and which might bear relevance to related exotic phenomena such as the complex phase diagram of URu$_2$Si$_2$ \cite{oh,mydosh,scheerer}.

\begin{acknowledgments}
We acknowledge the support of the LNCMI-CNRS, member of the European Magnetic Field Laboratory (EMFL). We gratefully acknowledge support by the Braunschweig International Graduate School of Metrology B-IGSM and the DFG Research Training Group GrK1952/1 ”Metrology for Complex Nanosystems”. Work at the National High Magnetic Field Laboratory was supported by the National Science Foundation Cooperative Agreement No. DMR-1157490 and the State of Florida, as well as the Strongly Correlated Magnets thrust of the US DoE Basic Energy Science “Science in 100T” program.
\end{acknowledgments}

\end{document}